# Matched Starts, Divergent Objects: Research-Object Formation in Human–AI Scholarly Work

Mehmed Zahid Çögenli, PhD
Department of Occupational Health and Safety, Faculty of Health Sciences
Uşak University, Uşak, Türkiye
Corresponding author: Mehmed Zahid Çögenli, mzahid.cogenli@usak.edu.tr
ORCID: 0000-0003-3018-4157

**Abstract**

Human–artificial intelligence interaction can shape what scholarly inquiry becomes about. This study examines research-object formation in two isolated scholarly trajectories developed by the same expert researcher with Claude and ChatGPT under matched starting conditions. Both used the same five-session, 843-turn project corpus, frozen research problem, initial prompt, publication objective, and conduct rules. The subsequent scholarly-development conversations were reconstructed ex post through consequential-event analysis, a Socioduality relational-process overlay, and tracing into completed manuscripts. Both trajectories shifted the continuity problem from recall toward usability, then developed different research objects: continuity labour and distributed, evolving project state. Failed analytical units, rejected explanations, counterexamples, and changes in analytical scale shaped these objects and propagated into research questions, methods, findings, evidence logic, and theoretical contributions. Consequential scholarly change, reciprocal relational continuity, and local substantive re-formation had distinct boundaries; continued interaction did not itself establish epistemic endorsement. The findings demonstrate empirically constrained research-object formation under matched scholarly starts. For systems supporting sustained scholarly knowledge work, they motivate recoverable analytical provenance, inspectable proposal status, and source-faithful contribution records.

## 1. Introduction

Scholarly knowledge is typically encountered in stabilized form: as research questions, methods, findings, concepts, and completed texts. These products compress the histories through which analytical possibilities became visible, alternatives were rejected, and the object of inquiry acquired form. Process research emphasizes that explaining emergence requires attention to sequence, activity, and how earlier developments become conditions for later ones (Langley et al. 2013). Studies of scientific practice similarly treat research as a knowledge-producing process in which problems, categories, evidentiary demands, and objects of inquiry are progressively articulated and revised (Knorr-Cetina 1981; Knorr Cetina 1999; Pickering 1995; Rheinberger 1997).

Sustained human–AI scholarly work makes this formation process an interaction problem. Researchers work with generative AI to interpret problems, select analytical approaches, test evidence, develop concepts, and construct manuscripts. AI proposals and reformulations can become consequential conditions for subsequent human judgment and action. Recent studies document the articulation, relationship, and identity work through which academics develop working relationships with generative AI, alongside changes in writing strategies and AI contributions across repeated activity (Boulus-Rødje et al. 2024; Dhillon et al. 2024; Yang et al. 2025). Human–AI interaction thus participates in scholarly development while leaving a detailed record through which that development can be reconstructed.

The distinction at stake is between producing different answers about a fixed object and forming different objects of inquiry: different phenomena around which questions, evidence, and explanations are organized. Failed analytical units, rejected explanations, counterexamples, and changes in analytic scale can alter what subsequently becomes salient and what conceptual account can be sustained. A completed manuscript shows what stabilized, but does not by itself explain how the inquiry came to take that form.

Many-analysts studies establish that the same data and overarching question can yield divergent results through different defensible analytical choices (Silberzahn et al. 2018; Landy et al. 2020; Breznau et al. 2022). Kummerfeld and Jones (2023) further show how an overarching question can become different actionable questions during analysis. Building on this evidence of divergence under shared starts, the present study examines research-object formation through recorded process histories and traces its consequences into complete scholarly products.

The study followed two isolated scholarly-development trajectories in which the same expert researcher worked separately with Claude and ChatGPT to develop complete manuscripts. Both trajectories began with Project P, an existing five-session, 843-turn corpus documenting a mobile-application development project, and the same frozen research problem concerning continuity in prolonged human–AI work. Project P was the shared material being investigated; the subsequent scholarly-development conversations were the process records analyzed here. Initial instructions, publication objective, and researcher conduct rules were also held common, while theory choice, analytical strategy, conceptual development, and follow-up interaction evolved within each trajectory. The common problem was selected through a preceding matched problem-generation phase, and formal comparison began only after the manuscripts were completed. A third branch was excluded when predefined input-equivalence and auditable full-corpus processing requirements remained unmet after remediation. The unit of comparison is the developing human–AI trajectory.

The central finding is empirically constrained research-object formation under matched scholarly starts. Both eligible trajectories shifted the continuity problem from recall toward usability, but developed different objects of explanation. In the Claude trajectory, failures of temporal and topical delimitation redirected analysis toward the distributed work required to sustain usable collaboration, developed as continuity labour. In the ChatGPT trajectory, a whole-case analytical reset redirected attention toward distributed, evolving, and unevenly usable project state. These objects acquired empirical content through distinct histories of testing, rejection, and conceptual stabilization. Their formation propagated into different research questions, finding architectures, methodological accounts, evidence logics, and substantive scholarly contributions.

### 1.1. Research question

How do human–AI scholarly trajectories diverge under matched starting conditions, and how do their process histories shape the research objects and scholarly contributions they ultimately produce?

The study makes three connected contributions. First, it provides a matched, traceable empirical account of how different research objects form from the same substantive problem and corpus through sustained human–AI interaction. Second, it reconstructs the analytical developments through which this differentiation occurred and propagated into complete manuscripts with distinct substantive contributions. Third, a Socioduality relational-process analysis distinguishes consequential scholarly change, reciprocal relational continuity, and local substantive re-formation. It shows that continued interaction does not necessarily imply epistemic endorsement: proceeding with an interaction need not mean adopting its substantive claims. These findings inform the design of systems for sustained human–AI knowledge work, motivating explicit endorsement states, recoverable project-state provenance, and source-faithful preservation of contribution boundaries. Together, the contributions explain how interaction histories shape what scholarly inquiry becomes about and identify what systems must preserve to make that development inspectable.

# 2. Theoretical Background

## 2.1. Scholarly knowledge production as a process

Process scholarship treats stabilized outcomes as temporally constituted. Explanation requires reconstructing how earlier events became conditions for subsequent action, how alternatives gained or lost viability, and how provisional arrangements acquired sufficient stability to shape what followed (Langley 1999; Langley et al. 2013). Applied to scholarship, this distinguishes the final research question, analytical model, findings, or manuscript from the process through which it was produced.

Studies of scientific practice locate knowledge production in situated arrangements through which problems become workable, categories are revised, and evidence acquires significance through analytical activity (Knorr-Cetina 1981; Knorr Cetina 1999; Pickering 1995). Conceptual, methodological, material, and practical elements develop in relation to one another. A completed account can therefore compress a history of trial, correction, rejection, and stabilization whose consequences remain embedded in the final product.

Many-analysts research demonstrates that common starting materials permit different defensible analytical choices and divergent conclusions (Silberzahn et al. 2018; Landy et al. 2020; Breznau et al. 2022). Kummerfeld and Jones (2023) show that variation can also arise as an overarching question becomes different actionable questions. These findings establish the importance of examining how analysis develops under shared starts. For the present study, the task is to connect differences between completed scholarly products to the recorded developments through which their questions, evidence, and explanations acquired form.

## 2.2. Historical conditioning and paths-in-the-making in scholarly development

Analytical work does not restart from an identical decision space at every step. Once a distinction has been introduced, a coding strategy challenged, an explanation rejected, or an evidentiary requirement imposed, subsequent inquiry proceeds under conditions partly created by those developments. History matters through the capacity of earlier events to modify what later becomes analytically available, plausible, or consequential.

Scholarly inquiry can remain revisable while acquiring a consequential history. Path-creation accounts are useful here because they treat paths as emerging through situated action, challenge, redirection, and temporary stabilization (Garud et al. 2010). This differs from explanations centered on self-reinforcement and eventual lock-in. It is also compatible with accounts of scientific practice in which agency and material resistance iteratively reshape one another as inquiry proceeds (Pickering 1995).

In scholarly work, a failed segmentation strategy may redirect attention toward phenomena that it obscured; a counterexample may force a change in analytic scale; and a whole-case reconstruction may reveal changes inaccessible under a narrower unit. Such developments can accumulate without any single event determining the endpoint. What matters is whether the record shows how a particular development altered subsequent analysis and whether its consequences remained visible downstream.

A process genealogy therefore needs to identify the events through which analytical possibilities were opened, revised, or abandoned, and connect those events to later conceptual and methodological developments. Historical conditioning becomes an empirical relation to reconstruct through the trajectory.

### 2.3. From analytical trajectory to research-object formation

Historical development can transform what an inquiry is understood to be about. Work on epistemic objects and scientific practice shows how objects acquire specificity, change analytical status, or are reconstituted as research proceeds (Rheinberger 1997; Knorr Cetina 1999; Daston 2000). Fujimura's (1987) account of constructing "do-able" research problems likewise locates their formation in the alignment of tasks, resources, and social organization. These perspectives make the constitution of the research object part of the knowledge-producing process.

Rheinberger's account of epistemic things is particularly useful because it treats inquiry as productive: an object can remain partly indeterminate and become progressively articulated through experimental and analytical work (Rheinberger 1997). Laboratory studies similarly show how scientific facts and objects acquire stability through situated knowledge-making practices (Knorr-Cetina 1981; Latour and Woolgar 1986). The object that organizes the completed account may thus depend on developments occurring during the inquiry itself.

Different analyses of the same material can organize it around different relations, distinctions, and explanatory problems. The resulting differentiation can extend into research questions, finding architectures, evidentiary demands, methodological accounts, and theoretical contributions. Research-object formation consequently concerns how these elements come to cohere around a particular phenomenon as the trajectory develops.

The empirical task is to reconstruct that formation from the development record. This requires tracing how proposals acquired substantive content, how empirical tests and counterexamples altered their scope, and how rejected alternatives affected subsequent inquiry. Under matched starting conditions, such reconstruction permits examination of how different objects formed from a common problem and corpus, and how the evidence constrained the accounts that could be sustained.

### 2.4. Human–AI interaction in scholarly knowledge work

Research on sustained human–AI work places interaction within evolving practices. In a four-month autoethnographic study, Boulus-Rødje et al. (2024) identify articulation, relationship, and identity work involved in academics' development and maintenance of working relationships with generative AI. Experimental co-writing research shows how different levels of AI scaffolding affect writing quality, productivity, and the experience of ownership (Dhillon et al. 2024). A diary study of LLM-assisted writing with speech documents how writers adapt the tool's functions to their own tasks and writing strategies (Yang et al. 2025). These studies provide complementary evidence about how human activity and AI assistance are organized in practice.

Situated accounts of human–machine relations provide a basis for examining how action is organized through engagement with computational systems and the circumstances of use (Suchman 2007). Research on AI integration likewise makes the human effort required to sustain use visible. Fox et al. (2023), studying frontline work with automation, identify calibration, troubleshooting, and repair as forms of "patchwork" that bridge gaps between technological promises and practical accomplishment.

These perspectives direct attention to the work through which AI contributions become usable within an ongoing activity.

The terminology of collaboration also requires analytical precision. Sarkar (2023) argues that the metaphor can obscure human labour and its attribution. In this article, human–AI collaboration describes the iterative scholarly work arrangement, with authorship and responsibility remaining with the researcher. The analysis examines the content and consequences of human and AI contributions as they develop through interaction, while keeping their roles and contribution boundaries identifiable.

For scholarly knowledge work, this means following how proposals, challenges, empirical checks, and reformulations alter the direction of inquiry. Human–AI interaction participates in research-object formation, and its preserved record permits reconstruction of that participation. The present study therefore examines developing scholarly trajectories, linking interaction histories to the objects and contributions stabilized in completed manuscripts. It also examines whether consequential scholarly change, reciprocal interactional continuity, and local substantive re-formation follow the same boundaries. This connects the process perspective to an empirical account of sustained human–AI interaction in the production of scholarly knowledge.

# 3. Methods

## 3.1. Research architecture

The study comprised three phases: common-problem generation (P0), isolated scholarly development (P1), and ex post comparative process analysis (P2). P0 established a shared substantive problem from the same empirical corpus. In P1, the same expert researcher developed separate scholarly trajectories with different generative AI systems under matched starting conditions. Subsequent analytical and interactional development proceeded naturally within each branch.

P2 began only after the eligible trajectories had reached completed manuscript form. It combined scholarly trajectory analysis (P2-A), which reconstructed consequential events and their downstream effects, with a Socioduality relational-process overlay (P2-B), which examined reciprocal relational continuity and local substantive re-formation in the same interaction records. Downstream propagation analysis connected these developments to the completed scholarly artifacts. This sequencing kept the formal comparison outside the development of the branch manuscripts. Fig. 1 summarizes the research architecture.

**Fig. 1** Research Architecture

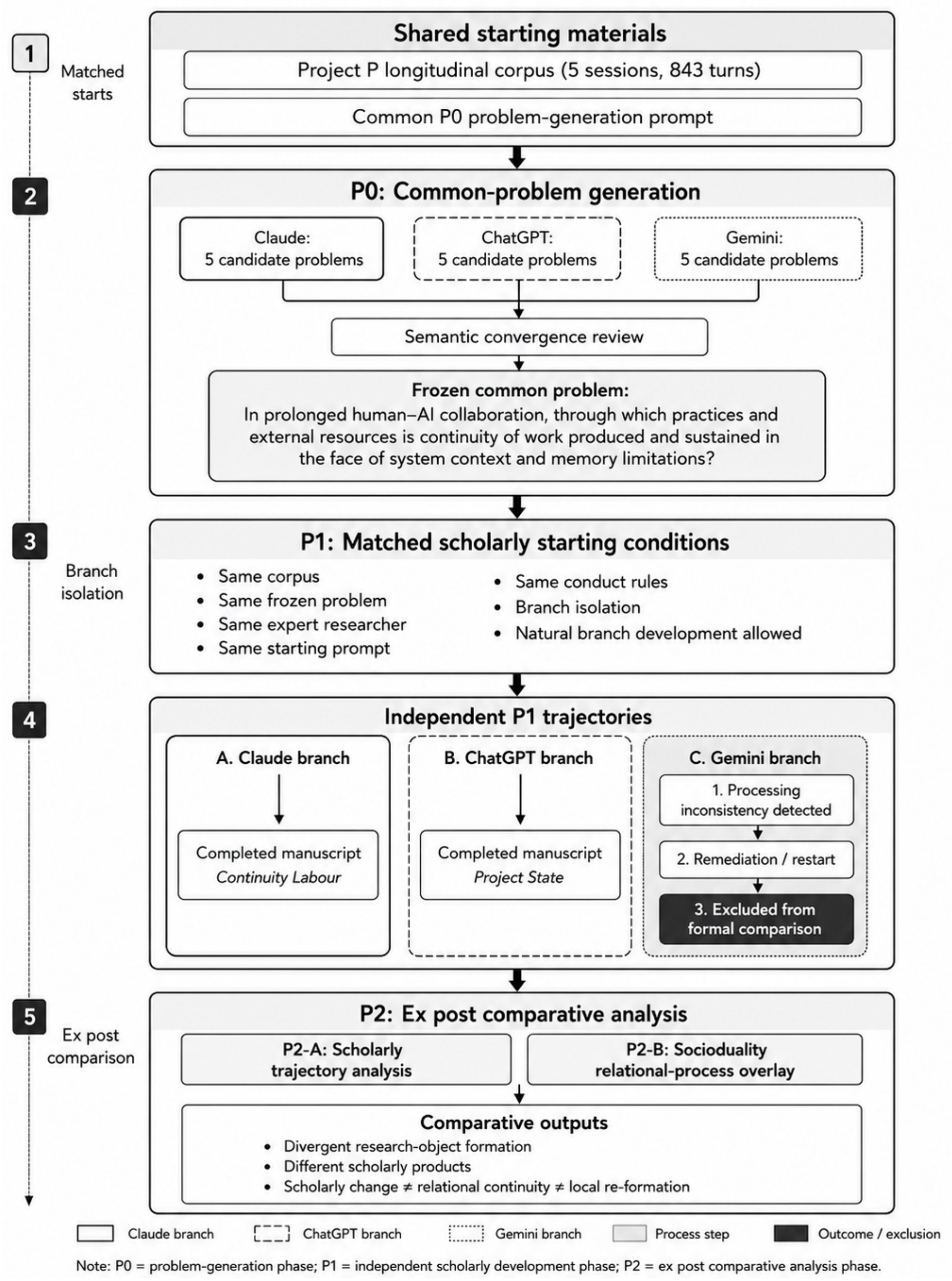

*Note. P0 = problem-generation phase; P1 = isolated scholarly-development phase; P2 = ex post comparative analysis phase. The figure uses an abbreviated formulation of the frozen common problem; its full wording is provided in Section 3.3.*

Alt text: Flowchart showing Project P as the shared input, followed by independent problem generation, selection of a frozen common problem, and isolated scholarly trajectories under matched starting conditions. Claude and ChatGPT produce completed manuscripts; Gemini is excluded following an input-integrity check. Ex post analysis compares scholarly development, relational processes, and propagation into the endpoint artifacts.

### 3.2. Project P: the natural longitudinal corpus

Project P was a naturally occurring longitudinal human–AI interaction corpus documenting the conceptualization and development of a real mobile-application project involving a non-technical founder and an AI assistant. It comprised five conversational sessions and 843 turns: 288 in Session 1, 277 in Session 2, 160 in Session 3, 70 in Session 4, and 48 in Session 5.

The corpus existed before the comparative study and recorded an ongoing work process. Its activities included problem clarification, design decisions, specification work, constraint management, continuity restoration, revision, and the use of externalized project materials. It was anonymized before P0 and P1, and all eligible branches received the same version. This fixed input preserved the complexity of the longitudinal project while removing corpus selection as a source of between-branch variation.

Two levels of empirical material must be distinguished. Project P supplied the shared material investigated in the branch studies. The subsequent Claude and ChatGPT scholarly-development conversations supplied the process records analyzed in the present comparative article. The 843 turns therefore describe the common input corpus; the P1 records document how scholarly inquiry developed within each trajectory.

### 3.3. P0 common-problem selection

ChatGPT, Claude, and Gemini independently received the same Project P corpus and problem-generation prompt. Each system was asked to propose five research problems that were theoretically meaningful, examinable through the corpus, and sufficiently substantial to support a publishable academic article. The systems were not informed of the other branches or the downstream comparison.

The resulting 15 candidates were compared for semantic convergence: whether they addressed the same underlying phenomenon despite differences in formulation. Data fit, originality, and publishability served as secondary criteria where more than one common direction was identifiable. Table 1 in Supplementary File 3 presents the candidates in their original within-system order and their primary semantic clustering.

The selected common direction concerned continuity in prolonged human–AI collaboration under system context and memory limitations. It was frozen for P1 as:

In prolonged human–AI collaboration, through which practices and external resources is continuity of work produced and sustained by the user in the face of the system's context and memory limitations?

P0 established the common starting problem before the scholarly trajectories developed. It served as a design control, while the substantive comparison examined the two eligible P1 trajectories. The frozen P0 and P1 prompts are provided in Supplementary File 3.

### 3.4. P1 matched conditions and natural scholarly trajectories

The matched conditions were the anonymized Project P corpus, frozen research problem, expert researcher, initial scholarly-development prompt, publication objective, and researcher conduct rules. The two eligible trajectories used ChatGPT (GPT-5.6 Sol) and Claude Opus 5 (High). These study-recorded user-facing configurations constituted the intentional branch-defining difference at P1 entry.

The initial prompt specified a complete, publishable research article at a level suitable for at least a Scopus Q2 journal. It left theory selection, research questions, method, coding strategy, analytical procedure, and conceptual contribution to be developed through interaction. The systems were instructed to begin by evaluating the corpus and problem, proposing how the research should proceed, and explaining that proposal, rather than completing the task in one response. Working discussion and decision communication took place in Turkish; manuscript titles, abstracts, keywords, final research questions, and sections were developed in academic English.

Researcher conduct was governed by common epistemic principles: fidelity to the full corpus, preservation of the project-level context, empirical support for claims, scrutiny of attractive explanations, resistance to premature closure, and revision or abandonment of unsupported interpretations. Follow-up utterances responded to each branch's developing analytical state. Analytical units, strategies, theoretical choices, conceptual developments, empirical checks, manuscript architecture, and revision pathways were consequently trajectory-emergent. The design matched the conditions governing participation while preserving the interaction through which scholarly work developed.

The branches remained isolated throughout P1. Neither had access to the other's interaction record, manuscript, concepts, procedures, or findings, and deliberate transfer of branch-specific outputs, methods, sources, analytical solutions, or decisions was prohibited. Formal comparison began after completion. Table 1 distinguishes the matched conditions, branch-defining AI-system difference, and trajectory-emergent elements.

**Table 1.** Matched, Branch-Defining, and Trajectory-Emergent Conditions

| **Design element** | **Cross-branch status** | **Operationalization** |
|---|---|---|
| Empirical material | Matched | Same anonymized Project P corpus |
| Substantive problem | Matched | Same frozen common problem |
| Expert researcher | Matched | Same expert researcher |
| Initial P1 prompt | Matched | Same scholarly-development prompt |
| Publication objective | Matched | Same complete publishable-article objective |
| Researcher conduct rules | Matched | Same corpus-fidelity, integrity, and evidentiary principles |
| AI system/model configuration | Intentionally different | Eligible P1 trajectories: ChatGPT (GPT-5.6 Sol) and Claude Opus 5 (High) |
| Follow-up interaction | Trajectory-emergent | Branch-specific interactional development |
| Analytical unit | Trajectory-emergent | Branch-specific |
| Theory selection | Trajectory-emergent | Branch-specific |
| Coding / analytic strategy | Trajectory-emergent | Branch-specific |
| Conceptual development | Trajectory-emergent | Branch-specific |
| Validation procedure | Matched standard; emergent procedure | Same expectation of corpus fidelity and claim support; branch-specific tests, challenges, and audits |
| Manuscript architecture and revision pathway | Trajectory-emergent | Branch-specific |
| Cross-branch transfer | Prohibited | No deliberate transfer of P1 outputs, concepts, methods, sources, analytical solutions, or decisions |

*Note. The design standardized the conditions governing the researcher, not the researcher's utterances. The AI system/model was the intentional branch-defining difference; subsequent scholarly development was not scripted.*

### 3.5. Gemini integrity control and exclusion from formal comparison

Gemini entered P1 under the same intended starting conditions. Repeated inconsistencies in its handling of Project P included undercounting the five-session, 843-turn corpus and uncertainty about whether analyses used the complete input. Input equivalence and auditable full-corpus processing were prerequisites for formal comparison.

A remediation procedure restarted the branch using Gemini 3.1 Pro in its advanced reasoning mode, with renewed emphasis on full-corpus access, structure, and processing requirements. Subsequent claims of complete analysis remained insufficiently reconcilable with the observable record to establish these prerequisites. The branch was therefore excluded before P2 under the study's predefined integrity requirements.

Gemini remained part of the design and integrity record, including P0 participation, P1 initiation, remediation, and exclusion. It was excluded from trajectory, relational-process, and endpoint comparisons. The decision concerned eligibility for an auditable matched-input comparison and was not a judgment of the substantive quality or direction of its emerging findings.

### 3.6. Comparative unit and endpoint scholarly artifacts

The unit of comparison was the human–AI scholarly trajectory: the temporally ordered process through which the researcher and an AI system developed a substantive article from the common P1 conditions. It included proposals, challenges, decisions, rejected alternatives, empirical checks, conceptual developments, and revisions. Each system contributed one eligible trajectory, so system-level effects cannot be separated from trajectory-level effects. Findings are reported as properties of the observed scholarly processes.

Both trajectories culminated in substantive scholarly manuscripts whose questions, methods, findings, evidence structures, and contributions were developed and evaluated through sustained interaction. The Claude endpoint was Continuity Labour: Sustaining Prolonged Collaboration with a System That Does Not Remember. The ChatGPT endpoint was Project State in Long-Term Human–AI Collaboration: Distributed, Evolving, and Unevenly Available. Frozen copies are provided as Supplementary Files 1 and 2, respectively.

Each manuscript was frozen before P2 and served two analytical roles: documenting the scholarly product in which its trajectory stabilized and providing downstream artifact traces for earlier developments. Consequential events were identified from the process records. The endpoint manuscripts were then examined to establish whether those events' consequences were retained, transformed, or abandoned. Event selection therefore preceded assessment of endpoint propagation.

### 3.7. P2-A: scholarly trajectory analysis

P2-A reconstructed consequential scholarly development within each eligible branch. An event was included when it changed, constrained, stabilized, redirected, or abandoned a scientific choice with consequences for later work. Routine continuation, stylistic editing, ordinary task completion, and exchanges without consequential downstream effects were excluded.

Candidate events were examined through the following process grammar:

trigger → human move → AI move → interactional resolution → scholarly consequence → downstream propagation → artifact trace

An event was eligible when it changed or locked a scientific decision, redirected the analytical strategy, introduced a recurring concept or procedure, narrowed or rejected a claim under contradictory evidence, altered the scale or object of analysis, or materially affected the final scholarly artifact. Abandoned routes remained analytically relevant when their rejection changed subsequent inquiry. Failed segmentation schemes, rejected analytical units, unsupported explanations, counterexamples, and methodological reversals were therefore retained where they had traceable consequences.

Events were reconstructed separately for the two branches, with no requirement that they contain equivalent numbers or types of turning points. The Claude ledger contained 16 events (C1–C16), including eight principal turning points used in the cross-branch trajectory map: C2, C3, C5, C8, C10, C11, C13, and C15. The ChatGPT ledger contained 18 events (G1–G18). The ledgers represented each trajectory's consequential history; cross-branch comparison examined their developmental patterns rather than imposing a common event sequence.

The comparison examined where scholarly redirection occurred, which alternatives were retained or abandoned, how earlier decisions altered later analytical conditions, and how their consequences propagated into completed scholarly products. This connected endpoint differences to the analytical genealogies documented in the interaction records.

### 3.8. Source fidelity and move-reconstruction audit

Before relational coding, each P1 transcript underwent a source-fidelity audit. A move was defined as a separately recorded focal contribution by one party. Technical or display segments belonging to one observable assistant response were merged into that response contribution. Moves were not inferred from timestamps, telemetry, hidden system behaviour, or assumed intent. The observable contribution structure governed unitization.

In the ChatGPT branch, an initial machine-readable representation treated technical assistant-side splits as separate nodes. It contained 305 apparent nodes, including 91 same-party AI-to-AI adjacencies, producing artificial fragmentation of the relational topology. Reconstruction against the source transcript yielded 214 moves: 107 human and 107 AI, with no same-party adjacency. The 213 observable adjacent cross-party boundaries were unchanged; the correction concerned the representation of compound responses and the topology inferred from it.

The Claude transcript was assessed under the same rules. Its representation already corresponded to the observable alternating move structure and required no comparable correction. All candidate construction, boundary assessment, episode coding, and pathway identification used the source-faithful ledgers. The audit established the interaction units before relational structure was inferred from the exported records.

### 3.9. P2-B: Socioduality relational-process overlay

P2-B applied Socioduality as a separately specified relational-process overlay to the source-faithful P1 records. Its purpose was to examine whether consequential scholarly development coincided with reciprocal relational continuity and local substantive re-formation (RF). Operational coding followed the frozen Socioduality Coding Manual v0.2; construct meaning and boundaries followed the active Socioduality specification (Çögenli 2026).

A sociodual process was treated as a sequential, history-carrying reciprocal relation between two distinguishable parties, in which one party's response becomes part of the observable conditions for the other's subsequent move. Every contiguous alternating three-move sequence, H–A–H or A–H–A (H =

human; A = AI), was identified as a candidate. The two adjacent cross-party boundaries were assessed separately before candidate-level episode status was derived.

Boundary status was Supported when evidence for relational dependence was stronger than temporal succession alone; Absent when the record positively indicated that the later move was independent of the preceding contribution; and Insufficient when dependence was compatible with the record but could not be established with sufficient evidence. A candidate was Confirmed only when both boundaries were Supported. Any Absent boundary produced a Non-sociodual candidate. Candidates with no Absent boundary but at least one Insufficient boundary were Indeterminate.

Secondary coding of orientation and substantive re-formation was restricted to Confirmed episodes. Substantive re-formation compared the third move with the first. It was Present when an outcome-relevant substantive change was observable, Absent when the third move remained substantively comparable despite continuation, acceptance, rejection, reaffirmation, verification, or surface reformulation, and Insufficient when a defensible comparison was unavailable. Reciprocal continuation and substantive change were therefore assessed separately.

Overlapping Confirmed episodes were joined into maximal uninterrupted pathways. A pathway required at least two overlapping Confirmed episodes; Non-sociodual or Indeterminate candidates terminated it. A single Confirmed episode did not constitute a pathway. Pathway length represented relational continuity, while re-formation rates described the observed trajectories. Neither measure was interpreted as a score for epistemic superiority, collaboration quality, or model performance.

The resulting relational structure was overlaid on the separately reconstructed P2-A event structure to examine whether scholarly turning points, pathway boundaries, and local substantive re-formations occupied the same temporal locations.

### 3.10. Downstream propagation analysis

Propagation analysis traced how consequential P2-A developments affected later inquiry and the completed manuscripts. Each event was followed through the subsequent interaction record and endpoint artifact across six domains:

research object → research question → finding architecture → method narrative → evidence logic → final scholarly contribution

Developments were classified by their downstream fate. Retained developments remained substantively recognizable in the final product. Transformed developments persisted with changes in analytical form, scope, terminology, or theoretical role. Abandoned developments were removed following empirical challenge, counterexample, methodological failure, or later reconstruction.

Abandonment remained consequential when rejecting a route redirected subsequent inquiry. Its role could therefore be traced even when the rejected account was absent from the endpoint manuscript. Propagation required an observable sequential connection between an earlier development and later scholarly decisions or artifact structure; similar terminology at different points was insufficient on its own.

This analysis connected differences between the final research objects, questions, findings, methods, evidence structures, and contributions to identifiable developments within the trajectories. It also distinguished provisional analytical activity from developments retained or transformed in the structure of the completed scholarly product.

### 3.11. Researcher position and participation

The same expert researcher participated in both eligible P1 trajectories and conducted the P2 analysis. A common researcher reduced between-person variation in disciplinary expertise, corpus familiarity, publication standards, and substantive judgment. The researcher also actively questioned, evaluated, redirected, accepted, and rejected proposals. These interventions formed part of the scholarly processes under investigation.

Common conduct rules established a stable epistemic criterion across trajectory-specific problems. The form of intervention varied with each branch's analytical state, while the requirements for whole-project integrity and empirical defensibility remained common. The preserved records allowed researcher interventions to be examined alongside AI contributions in reconstructing consequential events and their propagation.

The Claude trajectory was completed before the ChatGPT trajectory. The researcher retained disciplinary expertise and procedural experience between branches; branch isolation prohibited deliberate transfer of branch-specific outputs, concepts, methods, sources, analytical solutions, and decisions. The design therefore concerns separately developed trajectories under a common researcher, with within-person carryover acknowledged. The same researcher was both a participant in scholarly development and the analyst reconstructing its process, making explicit criteria and preserved interaction records central to analytical transparency. Generative AI also supported language editing and stylistic refinement of this comparative manuscript; the author verified the resulting text and retains full responsibility for it.

## 4. Results

### 4.1. Matched starts, divergent analytic trajectories

The two eligible trajectories developed different analytical organizations under matched starting conditions. Both worked from the same Project P corpus and frozen problem, with the same researcher, initial prompt, publication objective, and conduct rules. Their analytical organization diverged during development, before the manuscripts were completed.

The Claude trajectory repeatedly proposed analytical structures, tested their boundaries against the corpus, and revised or abandoned those that could not be sustained. Bounded episode segmentation, a developmental explanation, and temporal and topical decompositions were successively challenged. The resulting inquiry increasingly organized the continuity problem around work embedded in ordinary project activity.

The ChatGPT trajectory developed through reconstruction and changes in analytical scale. An initially rich mechanism-level account was challenged when detailed local analysis displaced attention to the whole project. A whole-case reset subsequently made project change outside AI participation central to the explanation and refined distinctions between previously shared and never-observed state, and between availability and usability.

Researcher interventions reflected these different analytical states under the common requirement for whole-project integrity and empirical defensibility. In the Claude trajectory, scrutiny focused on boundaries that fragmented continuity-related activity. In the ChatGPT trajectory, it focused on selected mechanisms that had become more elaborated than the reconstruction of the case as a whole. The observed patterns were broadly separate → classify → test → eliminate or stabilize in the former, and reconstruct → compare → deepen → reset → distinguish → reframe in the latter.

Table 2 summarizes the shared convergence and subsequent divergence in research objects, analytical organization, and scholarly products.

**Table 2.** Shared Convergence and Divergent Scholarly Trajectories

| Dimension | Claude trajectory | ChatGPT trajectory |
|---|---|---|
| Shared analytical hinge | Recall → usability | Recall → usability |
| Dominant analytic organization | Separate → classify → test → eliminate/stabilize | Reconstruct → compare → deepen → reset → distinguish → reframe |
| Critical redirection | Segmentation and decomposition fail | Whole-case reset after micro-analytic overdevelopment |
| What becomes analytically central | Distributed work required to sustain continuity | Changing project state whose usability must be sustained |
| Research object | Continuity labour | Project state |
| Research-question architecture | Four questions on practices, configuration over time, distribution, and constraints | One reconstructed process question on distributed, evolving, and unevenly available state |
| Finding architecture | Practice repertoire + occasioning + co-production + platform constraints | State conditions + usability + risk + cumulative change |
| Evidence logic | Boundary testing, falsification, blind challenge, repertoire correction | Whole-case reconstruction, maximum variation, missing-piece analysis, longitudinal counterexamples |
| Completed scholarly product | Continuity Labour: Sustaining Prolonged Collaboration with a System That Does Not Remember | Project State in Long-Term Human–AI Collaboration: Distributed, Evolving, and Unevenly Available |

*Note. The shared hinge is shown explicitly because convergence preceded divergence in both branches.*

### 4.2. Convergence before divergence: from recall to usability

Both trajectories moved the initial continuity problem beyond recall toward usability. The central issue became whether relevant project state could support the work that followed. This shared shift connected the eventual objects to the same substantive problem.

In the Claude trajectory, unsuccessful attempts to isolate continuity work redirected attention toward the activity required to keep collaborative state usable. Documentation, state preservation, re-entry, and constraint maintenance became consequential as work distributed through the project. In the ChatGPT trajectory, reconstruction showed that previously shared information could remain available yet be stale or inappropriate to the current situation, while consequential state produced outside AI participation could have no prior AI representation to recover.

The ChatGPT trajectory also tested an elapsed-time explanation. A return after approximately 50 days proceeded without the expected continuity failure, while difficulties occurred under other conditions without comparably long interruptions. The account consequently differentiated usability risk from temporal distance, including both reactive restoration and anticipatory preparation for later use.

From this shared usability problem, the objects differentiated. Claude foregrounded the work required to maintain usability, developed as continuity labour. ChatGPT foregrounded the distributed and changing project state whose usability had to be maintained. The common analytical hinge was recall → usability; the subsequent formation concerned what that usability problem required explanation of.

### 4.3. Analytic genealogy and research-object formation

The two research objects acquired empirical content through distinct sequences of analytical proposal, challenge, testing, and revision. Their genealogies show how developments within the interaction changed the organization of inquiry. Quoted Turkish passages below are translated into English; move identifiers refer to the source-faithful P1 ledgers.

#### 4.3.1. Claude: from failed delimitation to distributed continuity labour

The initial strategy sought bounded continuity episodes within Project P. The researcher challenged the effect of this segmentation on the process being investigated:

Researcher (M009): "The segmentation is not wrong, but it seems to cut through the process."

Claude (M010): "53% of the decision-fixing acts fall outside the episode boundaries (68/128). … I am removing episodes as the primary unit of analysis."

The response connected the objection to an empirical check and a methodological change. Sensitivity testing across merge thresholds from 3 to 30 turns increased coverage smoothly from 21% to 62%, without a natural breakpoint. Bounded episodes were abandoned as the primary unit and retained as illustrative waypoints. The consequence was a change in where continuity-related work could be sought: actions excluded by the boundaries became evidence about the phenomenon's distribution.

The subsequent developmental interpretation also changed under testing. An initially proposed sequence of practices was undermined when several practices appeared together, and an account of declining repair did not withstand comparison under session-length controls. The resulting interpretation treated practices as occasioned by project conditions. Attempts to recover separability through temporal and topical decomposition encountered further empirical resistance.

Topical analysis identified 25 work domains across 351 domain-bearing turns. Of these turns, 59% addressed at least two domains and 127 addressed three or more. The mean was 2.40 domains per domain-bearing turn, with a maximum of 12; 85 of 300 possible domain pairs co-occurred at least five times. These results showed extensive coupling between continuity-related work and ordinary project activity.

The cumulative consequence was the formation of an object centered on the distributed work required to sustain usable collaboration. Continuity-related actions became visible within documentation, state preservation, re-entry, constraint maintenance, canonisation, deliberate reset, and other project activities. Their empirical organization supported the continuity-labour account. The endpoint manuscript preserved the segmentation result in its methodological account and the temporal and topical entanglement in its explanation of why this labour remained difficult to isolate.

The genealogy can be summarized as:

segmentation failure → developmental account revised → temporal and topical separation failed → distributed and embedded continuity work became analytically central → continuity labour

Each revision changed the conditions of subsequent inquiry. Failed delimitation redirected analysis toward work that the earlier units had obscured, and that redirection remained visible in the completed scholarly artifact.

### 4.3.2. ChatGPT: from mechanism-level depth to evolving project state

The ChatGPT trajectory initially developed a detailed account of continuity infrastructure, supported by corpus inventory, artifact tracing, and tests of candidate mechanisms. The decisive redirection occurred when the researcher questioned whether increasing depth was displacing attention to the whole case:

Researcher (M035): “How far has continually going deeper taken us away from the whole, and how does this affect the entire analysis?”

ChatGPT (M036): “Let us freeze all the analyses so far as ‘Analytic Memos / Candidate Mechanisms’ … We will go through all 843 messages chronologically.”

Earlier analyses were retained as candidates while chronological reconstruction became the primary task. The subsequent passes reconstructed project phases, artifacts, state changes, and later consequences before reassessing the mechanisms against the wider case. This changed the evidence relevant to the explanation: activities outside the AI conversation became consequential to understanding continuity within it.

During this reconstruction, the branch formulated the distinction explicitly:

ChatGPT (M052): “project state changes outside the AI interaction and must subsequently be brought back into the collaborative workspace.”

The emerging account differentiated recovery of previously shared state from incorporation of consequential state that the AI had never observed. Domain purchases, trademark registration, manual work, and later external evidence could change the project independently of AI participation. Such state could not be explained simply as information the system had forgotten.

Further testing differentiated availability from usability. A representation could be accessible but stale, or valid information could be applied to the wrong current situation. The long-gap counterexample described in Section 4.2 also constrained an elapsed-time explanation. Together, these developments organized the inquiry around what state existed, where it existed, whether it had been observed, whether it remained current, and whether it could support subsequent action.

The endpoint manuscript's research question consequently centered on continuity when relevant project state was distributed, evolving, and unevenly available. Its findings distinguished the origins of continuity gaps, availability from usability, usability risk from elapsed time, and cumulative change without immediate supersession. The earlier reset thus propagated into both the object of explanation and the manuscript's intellectual architecture.

The genealogy can be summarized as:

mechanism-level overdevelopment → whole-case reset → project change outside AI participation became analytically consequential → previously shared versus never-observed state → availability ≠ usability → evolving project state

### 4.3.3. Comparative consequence: process histories formed different objects

The two genealogies show how a common substantive problem became organized around different explanatory objects. One trajectory formed an account of the work required to sustain usable collaboration; the other formed an account of the changing project state whose usability had to be sustained. Empirical checks and rejected alternatives shaped the content and boundaries of both accounts.

This differentiation propagated into the endpoint questions and finding structures. The Claude manuscript organized four questions around continuity practices, their configuration, the distribution of labour, and platform constraints. The ChatGPT manuscript organized one process question around distributed, evolving, and unevenly available state. These architectures carried the consequences of the preceding analytical histories.

Earlier events changed what each trajectory could subsequently ask, test, and sustain as an explanation. Research-object formation was therefore historically conditioned and empirically constrained. Fig. 2 presents the selected genealogies, their shared convergence, and the subsequent process-structure overlay.

**Fig. 2** Analytic Genealogies and Process Structures

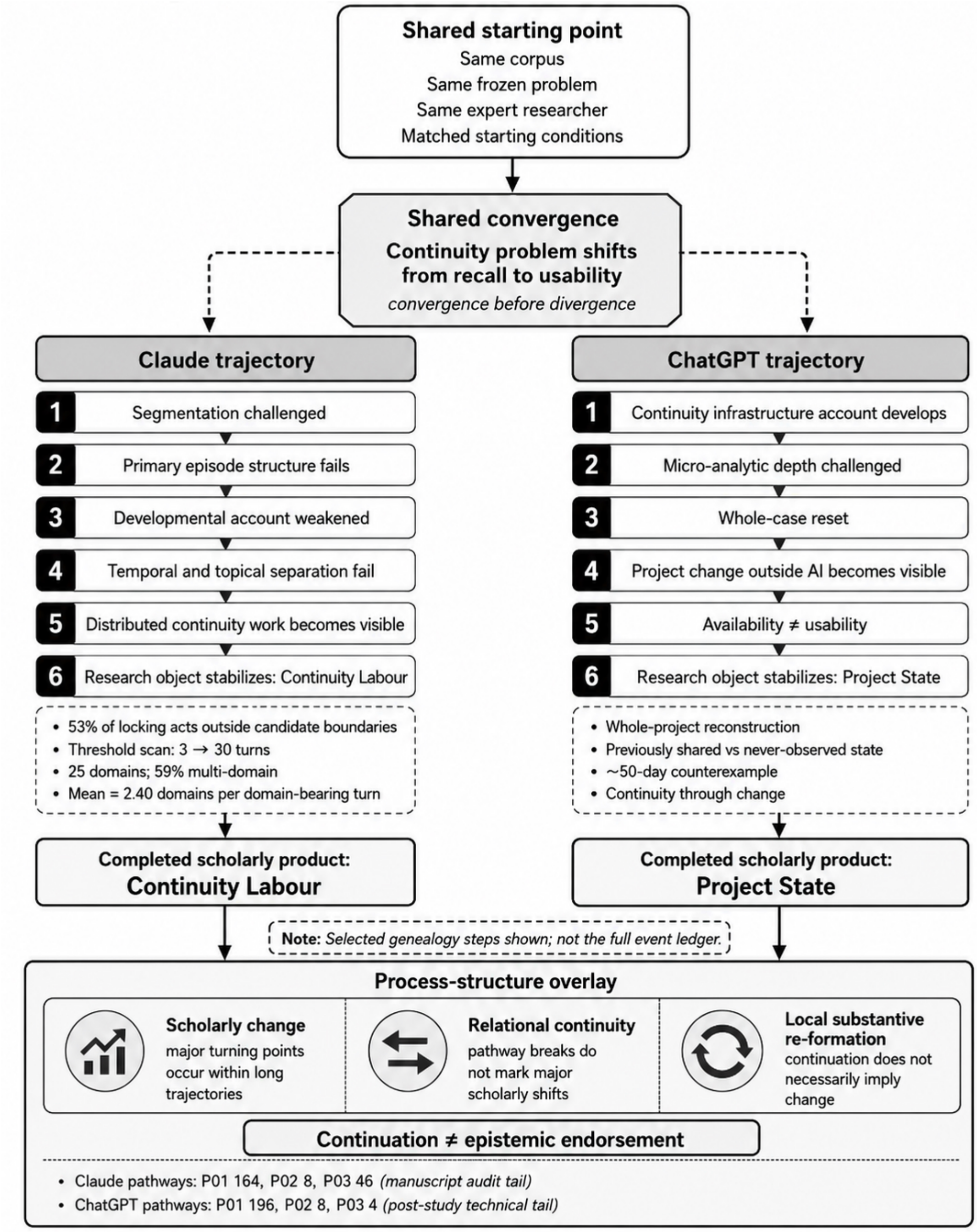

Alt text: Diagram showing a shared shift from recall to usability and subsequent divergence into continuity labour in the Claude trajectory and project state in the ChatGPT trajectory. Analytical challenges and revisions lead to the two endpoint manuscripts. An overlay distinguishes scholarly change, relational continuity, and local substantive re-formation, showing that continued interaction does not itself establish epistemic endorsement.

The next analysis examines whether these consequential scholarly changes coincided with boundaries in reciprocal relational continuity and local substantive re-formation.

### 4.4. Scholarly change, relational continuity, and local re-formation were distinct process structures

The P2-A scholarly-event reconstruction and the Socioduality overlay identified three process structures with different boundaries: consequential scholarly change, reciprocal relational continuity, and local substantive re-formation. In both trajectories, major analytical redirections occurred within long uninterrupted reciprocal pathways. The few pathway breaks, conversely, did not mark major scholarly turning points.

In the Claude trajectory, 224 moves generated 222 candidate three-move episodes: 218 Confirmed (98.2%), two Non-sociodual (0.9%), and two Indeterminate (0.9%). The Confirmed episodes formed three maximal pathways. P01 extended from M001 to M166 and contained 164 linked Confirmed episodes; P02 contained eight and P03 contained 46. Re-formation was Present in 190 of the 218 Confirmed episodes (87.2%), Absent in 21 (9.6%), and Insufficient in seven (3.2%).

Seven of the eight core scholarly events occurred within P01. These included the failed segmentation and explanatory tests described in Section 4.3.1, alongside validity checking, schema revision, and literature recalibration. Across the regions corresponding to all eight core events, 38 of 39 Confirmed episodes were RF-Present. Substantial local change thus accompanied consequential scholarly development while reciprocal continuity remained intact.

The boundary that ended Claude's P01 had a different significance. B166, from AI M166 to Human M167, was coded Absent because the researcher initiated a platform/session-limit issue independently of the preceding contribution. This was a relational break without a major scholarly redirection. B176 was coded Insufficient because the referent needed to establish dependence could not be recovered from the clean transcript.

In the ChatGPT trajectory, 214 moves yielded 213 adjacent cross-party boundaries and 212 candidate episodes: 208 Confirmed (98.1%), two Indeterminate, and two Non-sociodual. The Confirmed episodes again formed three maximal pathways. P01 extended from M001 to M198 and contained 196 linked Confirmed episodes; P02 contained eight and P03 contained four. Re-formation was Present in 157 of the 208 Confirmed episodes (75.5%) and Absent in 51 (24.5%); none were Insufficient.

Nearly all substantive scholarly development occurred within ChatGPT's P01, including the whole-case reset, recognition of project change outside AI participation, availability–usability distinction, research-question reconstruction, novelty recalibration, and stabilization of the findings and manuscript. P02 primarily involved late evidence restoration and manuscript refinement. B198, between title-lock confirmation and a new quotation/evidence audit, was coded Insufficient: continued dependence was compatible with the record, but explicit uptake evidence was insufficient. B208, from AI M208 to Human M209, was coded Absent because the researcher initiated a self-contained JavaScript transcript-extraction task after the manuscript audit. This introduced the functionally independent export task preceding P03 and did not redirect the manuscript's scholarly development.

Local re-formation differentiated the process further. Within a Confirmed episode, RF coding compared the third move with the first for outcome-relevant substantive change. Reciprocal dependence could therefore continue while a party's substantive position remained unchanged. In the ChatGPT trajectory, 48 of 104 human returns within Confirmed episodes were RF-Absent. Relational continuity alone did not establish local substantive change.

#### 4.4.1. Continuation did not equal epistemic endorsement

Transcript review distinguished permission to continue an analytical route from adoption of its claims. Human responses such as "continue," "let's proceed," and "let's see" could sustain further investigation without stating acceptance of the preceding explanation. RF coding described local substantive change; the interpretation of endorsement required attention to what the participant actually expressed.

One exchange illustrates this distinction. After narrowing a provisional explanation, ChatGPT proposed comparing alternative accounts against the corpus rather than treating the current account as settled (M056). The researcher replied:

Researcher (M057): "I am very curious to see where this will go. Let's continue."

The following AI contribution compared five candidate explanations: memory offloading, boundary reconstruction, distributed project state, ongoing state maintenance, and evolving partnership (M058). It retained the emerging account as provisional and proposed seeking counterexamples. The human move authorized further comparative analysis and expressed curiosity about its outcome; it did not state acceptance of the current explanation. Continued interaction therefore provided conditions for further scholarly work while epistemic commitment remained open in the expressed exchange.

This distinction also guided interpretation of the aggregate RF results. The differing RF-Present rates described how frequently local substantive change occurred within Confirmed episodes. They did not measure product quality, oversight, influence, or researcher passivity. Likewise, RF-Absent did not itself establish non-endorsement: an unchanged position and an unendorsed claim are different analytical judgments.

Two alternative explanations were examined. Technical or artifact-oriented tasks did not sufficiently account for the human RF-Absent episodes in the ChatGPT trajectory; many occurred during substantive analysis and involved permission to proceed. Runs of RF-Absent episodes also did not support a sufficiently consistent relationship with imminent scholarly redirection to retain that interpretation. The supported result remained local: continued work did not require continuous substantive re-formation, and continuation alone did not establish epistemic adoption.

The source-fidelity audit supplied a related methodological finding. Before the move reconstruction described in Section 3.8, technical assistant fragments produced an apparent topology of 19 pathways and 33 singletons. Reconstructing observable compound responses yielded the three-pathway structure reported here while preserving all 213 cross-party boundaries. Incorrect unitization therefore changed the apparent relational structure without changing the underlying cross-party exchanges.

Table 3 illustrates the non-equivalence of scholarly change, reciprocal continuity, and local re-formation. The endpoint differences consequently require examination of what changed within the continuing pathways and how those changes propagated into the manuscripts, addressed in Section 4.5.

**Table 3.** Non-equivalence of Scholarly Change, Relational Continuity, and Local Re-formation

| Trajectory | Illustrative event | Sociodual pathway position | Local re-formation | Relational break? | Downstream scholarly consequence |
|---|---|---|---|---|---|
| Claude | Failure of episode segmentation | Within P01 | Present | No | Primary candidate unit abandoned; distributed continuity work becomes visible |
| Claude | Temporal and topical decomposition fail | Within P01 | Present | No | Strengthens continuity-labour account |
| Claude | Platform/session-limit initiation | P01 boundary | Not coded | Yes | No major scholarly redirection |
| ChatGPT | Whole-case reset | Within P01 | Present | No | Analytic scale changes; project change outside AI becomes visible |
| ChatGPT | Availability ≠ usability distinction | Within P01 | Present | No | Propagates into research question and findings |
| ChatGPT | Transcript-export technical task | P02/P03 boundary | Not coded | Yes | No substantive manuscript-level redirection |

*Note. Local re-formation was coded only for Confirmed reciprocal episodes; "Not coded" therefore indicates structural inapplicability at pathway-breaking boundaries rather than missing data.*

### 4.5. Downstream propagation into the completed scholarly products

The developments reconstructed in Section 4.3 propagated into the endpoint manuscripts' research questions, findings, methods, evidence logic, and contributions. The artifacts documented both developments retained or transformed in the final accounts and the consequences of abandoning alternatives. Their architecture made the trajectories' different histories of inquiry visible.

#### 4.5.1. Propagation in the Claude trajectory

Failed delimitation became part of the Claude manuscript's method and substantive explanation. Bounded episodes were abandoned as the primary analytical unit after sensitivity testing failed to identify a defensible temporal boundary. The result survived in the endpoint's methodological account and its finding that continuity labour resisted temporal delimitation. Failed topical decomposition similarly became evidence that this work was embedded in activities whose immediate purpose lay elsewhere. Analytical failure thus acquired a different role in the completed account: it helped explain the labour's diffuse structure and practical invisibility.

The rejected developmental explanation also remained consequential. The endpoint explicitly reports that the proposed decline in repair did not survive control for session length. Its surviving finding specifies the practice repertoire as occasioned rather than developmental. The abandoned account therefore supplied a tested alternative against which the final explanation was defined.

The practice schema was retained after substantive revision. The external blind challenge used AI-generated readings of corpus excerpts under a protocol that withheld the developing practice schema. Category suggestions from these readings were checked against Project P before incorporation, expanding the repertoire from five to seven practices through the addition of canonisation and deliberate reset. The endpoint reports repair, externalisation, canonisation, deliberate reset, ritualisation, governance, and economisation. External input supplied challenges and candidate locations; the corpus supplied the evidence for the revised repertoire.

Literature recalibration changed the contribution's scope. An absence-based novelty claim was removed, while the account was specified around its analytical object, naturally occurring corpus, empirical organization, and temporal conditions. The completed manuscript consequently defines continuity labour as the effort required to keep previously established collaborative state usable across temporal discontinuities and explains how that effort occurs within ordinary project activity.

Four research questions organize the endpoint around practices and resources, their configuration over time, the division of labour between human and system, and platform-resource constraints. Its corresponding findings establish the seven-practice repertoire, occasioned configuration, interactional co-production, and the distinct burden associated with platform constraints. The additional delimitation finding connects the manuscript's explanation of invisibility to the analytical procedures that failed to isolate the work. Elimination, revision, and recalibration therefore shaped both the claims retained and the evidence supporting them.

### 4.5.2. Propagation in the ChatGPT trajectory

The whole-case reset changed the ChatGPT manuscript's analytical organization. Earlier conversation-centered and continuity-infrastructure accounts were abandoned as sufficient explanations. Chronological reconstruction became central to the method, while episodes served as comparative process units rather than an exhaustive partition of the corpus. The methodological account thus retained the shift in scale described in Section 4.3.2.

Distinctions developed through reconstruction organized the four findings. Finding 1 differentiates previously shared state that was no longer sufficiently usable from consequential state produced outside AI participation and never previously shared. Finding 2 separates availability from usability, including representational staleness and situational misapplication. Finding 3 locates continuity practices in realized or anticipated usability risk; the long-gap counterexample constrained an elapsed-time explanation. Finding 4 shows how continuity could support cumulative change without requiring new evidence to immediately supersede prior project state.

These developments also reorganized the research question:

How is work continuity constructed and sustained in long-term human–AI collaboration when relevant project state is distributed, evolving, and unevenly available across interactions, actors, and artifacts?

The question incorporated distinctions generated through the trajectory. The common starting problem's emphasis on practices and resources under context and memory limitations became an inquiry into continuity across changing project-state conditions.

The endpoint consolidated four findings into one overarching contribution: continuity is the ongoing accomplishment of keeping an evolving project sufficiently connected across previously shared, newly produced, unevenly usable, and changing states so that work can remain cumulative. Project state remained an operational term. The contribution resided in the process explanation and the distinctions supporting it. Whole-case reconstruction, comparative episodes, and counterexample testing consequently remained visible in the final research question, finding architecture, method, and contribution.

#### 4.5.3. From trajectory difference to scholarly-product difference

The completed products preserved different organizations of inquiry. Claude's four-question architecture examined continuity practices, their configuration, distribution, and constraints. ChatGPT's single process question examined continuity across distributed and evolving project state. Their evidence logic also differed: failed delimitation, developmental tests, and repertoire correction were integral to the former; whole-case reconstruction, contrasting episodes, and distinctions among state conditions organized the latter.

Propagation connected these differences across the scholarly layers specified in Section 3.10. What became the research object affected what was asked, which evidence was consequential, how findings were organized, and what contribution could be sustained. The endpoint differences were therefore traceable to developments within the interaction records, including rejected routes whose consequences survived in the final accounts.

### 4.6. Substantive scholarly contributions generated within the trajectories

Each trajectory produced a substantive account of the shared continuity problem. The Claude manuscript specifies continuity labour as a temporal form of articulation and gap-filling work under asymmetric persistence: state established through collaboration must repeatedly be made usable because it does not automatically remain available to both parties. Its account explains how this work is embedded in project activity, co-produced in interaction, and unevenly distributed. The system-authored re-entry protocol subsequently enacted by the participant provides one concrete expression of that co-production.

The ChatGPT manuscript specifies the project-state conditions that continuity work must address. Previously shared state may need reconstruction, externally produced state may need incorporation, and available state may require updating or correction of its application. Continuity can also require preparation for future usability risks and preservation of unresolved evidence alongside earlier commitments. The account explains what keeping an evolving project sufficiently connected entails when project activity and focal AI participation are intermittently coupled.

These contributions make different aspects of the common usability problem explanatory: the work required to sustain usable collaborative state and the changing state conditions that work must handle. Their distinct empirical findings and theoretical implications are documented in the frozen endpoint manuscripts. For the comparative study, the additional result is that their formation has recoverable process histories. Matched starts constrained the beginning of inquiry, while subsequent scholarly interaction formed different research objects and contribution structures, largely within continuous reciprocal pathways.

## 5. Discussion

### 5.1. Research-object formation under matched scholarly starts

The central contribution is empirically constrained research-object formation under matched scholarly starts. A common corpus and substantive problem became organized around continuity labour in one trajectory and project state in the other. The differentiation extended into research questions, methods, evidence logic, findings, and theoretical contributions. The preserved interaction records show how those scholarly structures acquired their form.

This finding connects with accounts of scientific practice in which research objects become articulated and stabilized through inquiry (Rheinberger 1997; Knorr Cetina 1999). The present study adds comparative process evidence: two separately developed human–AI scholarly trajectories began from matched conditions, and their developing objects can be followed through analytical tests, revisions, and completed manuscripts. Human–AI interaction participated in that formation by generating proposals, challenges, distinctions, and reformulations that became consequential to subsequent inquiry.

Empirical constraint was constitutive of the formation. In the Claude trajectory, failed delimitation helped specify the distributed and embedded character of continuity labour. In the ChatGPT trajectory, whole-case reconstruction made project change outside AI participation consequential to the account. Corpus testing and counterexamples changed what each explanation could sustain. The shared evidence permitted different explanatory organizations while constraining their content and boundaries.

The resulting objects also connect to established theoretical accounts. Continuity labour specifies temporal articulation and gap-filling under conditions in which previously established state must repeatedly be made usable (Strauss 1988). The project-state account develops the implications of cognitive organization across people, representations, and artifacts for an evolving project with intermittent AI participation (Hutchins 1995; Hollan et al. 2000). Their specific explanatory content developed through the recorded trajectories. The comparative contribution concerns how those different, empirically supported objects and contributions formed under common scholarly starting conditions.

### 5.2. From result variation to process explanation

Many-analysts research establishes that shared data, hypotheses, or overarching questions can support divergent analytical outcomes (Silberzahn et al. 2018; Landy et al. 2020; Breznau et al. 2022). Kummerfeld and Jones (2023) further distinguish an overarching question from the formal questions into which analysts translate it. This literature provides a metascientific basis for examining variation under common starting conditions.

The present study examines the development of complete scholarly accounts while keeping the expert researcher common. Its process records connect the starting problem to the research objects, questions, and contribution structures stabilized in the endpoint manuscripts. The comparison thus follows how analytical differences developed and became consequential across the scholarly trajectory.

The explanatory relation is sequential. A segmentation failure changed the relevance of subsequent decomposition strategies; a whole-case reset changed which project-state conditions entered the explanation; rejected accounts altered the alternatives available for later inquiry. An inventory of the decisions would identify their content, while their genealogy shows how one development changed the analytical conditions of those that followed. The additional explanatory layer is this historical conditioning of inquiry.

For longitudinal human–AI knowledge work, the trajectory consequently provides a unit that connects interaction to scholarly outcome. Local proposals and responses matter through the empirical tests, interventions, and downstream decisions to which they lead. This makes the relation between interaction history and completed knowledge products available for process explanation.

### 5.3. Historical conditioning and paths-in-the-making

The genealogies show how earlier analytical events changed what later became visible, plausible, or worth testing. Both trajectories nevertheless remained open to challenge, revision, and abandonment. This pattern is consistent with path creation, where situated action and temporary stabilization produce paths-in-the-making (Garud et al. 2010).

Researcher agency was part of that historical development. Challenges to segmentation, demands for whole-case reconstruction, and scrutiny of attractive explanations altered the course of inquiry. Each intervention occurred within an analytical situation partly constituted by preceding human and AI contributions. History mattered through the conditions it created for further action, including action that redirected the trajectory.

Downstream propagation makes this conditioning empirically identifiable. Consequential events affected later questions, evidence requirements, and contributions; rejected routes could persist as tested alternatives or as reasons for adopting a different analytical organization. Those connections were reconstructed across the process record and endpoint artifacts. The evidence therefore supports an account of revisable inquiry whose developing history shaped the objects and scholarly products it ultimately sustained.

### 5.4. Distinct process structures in longitudinal human–AI interaction

The Socioduality overlay distinguished reciprocal relational continuity from local substantive re-formation and the consequential scholarly changes reconstructed in P2-A. Approximately 98% of candidate episodes in both trajectories were Confirmed as reciprocally connected. Multiple changes in analytical unit, explanatory account, and research question occurred within these continuing pathways. The few relational breaks did not identify major scholarly transitions.

The overlay therefore added a relational structure to the scholarly genealogy. It showed where contributions remained reciprocally dependent while the inquiry changed, and where that dependence ended without scholarly redirection. Local re-formation supplied a further distinction: a connected return could leave one party's substantive position unchanged. These structures described different aspects of the same developing interaction.

Longitudinal HCI analysis consequently needs to specify the process relevant to its explanatory claim. A conversational boundary may locate a change in reciprocal dependence without identifying an intellectual turning point; an intellectual turning point may occur within an uninterrupted reciprocal pathway. Relational pathways, local substantive changes, and consequential scholarly events should be reconstructed at their respective levels and connected through the process record. Their separation makes it possible to examine how sustained interaction accommodates both analytical change and periods of substantive stability.

### 5.5. Continuation is not epistemic endorsement

The transcript analysis showed that permission to continue an inquiry and adoption of an explanation were different acts. The exchange in Section 4.4.1 authorized further comparison while the emerging account remained provisional. Other human returns likewise sustained analytical work without observable local substantive change. RF coding identified that change or stability; interpreting endorsement required examining the content and subsequent treatment of the proposal.

Provisional continuation has an epistemic function: allowing an account to develop can make its implications and weaknesses available for examination. Requests to proceed, positive transition language, or continued system use therefore cannot alone establish that a claim has been adopted. Assessments of AI influence and human oversight need to distinguish continued participation, substantive change, and epistemic endorsement, and follow how proposals are tested, revised, accepted, or abandoned downstream.

A design implication is to make proposal status inspectable in systems for scholarly knowledge work. Users should be able to mark an account as provisional, accepted, rejected, or still under test, and revise that status while retaining its rationale. A request to continue testing a candidate explanation should preserve its provisional status unless the researcher changes it. Such support would represent the distinction observed here between authorizing further analysis and committing to its claims.

### 5.6. Process history, endpoint evaluation, and system design

The endpoint manuscripts establish the scholarly products in which the trajectories stabilized. Their process histories explain how the objects, questions, and evidentiary structures became defensible. Reading the endpoints identifies continuity labour and project state as different accounts; reconstructing their development reveals the shared usability convergence, the consequences of failed analyses, and the analytical redirections within continuing reciprocal pathways. Linking product and history therefore adds explanation to the comparison of completed outputs.

This relation motivates systems that preserve recoverable analytical history alongside current manuscript content. A research question, conceptual distinction, or methodological choice should remain connected to the evidence and rationale through which it developed. An analytical reset should leave earlier candidate accounts and the reasons for reassessing them recoverable. Rejected segmentation, for example, should remain linked to the evidence of its failure and the analytical organization subsequently adopted. These links would support later scrutiny and revision of the scholarly decisions represented in the current artifact.

This requirement connects with work on the interpretation needed to make information usable across contexts (Bannon and Bødker 1997) and with design-rationale systems that link artifact changes to their discussion context (Li et al. 2025). For human–AI scholarly work, the relevant history includes provenance, empirical challenges, status changes, unresolved alternatives, and consequential abandoned routes. Selective recovery of this history would allow the researcher to inspect why an account changed and which evidence supported the surviving formulation.

The work required to maintain that history also deserves representation. Invisible-work research shows how visibility depends on the indicators through which work is represented (Star and Strauss 1999), while AI-integration research documents continuing human effort in calibration, troubleshooting, and repair (Fox et al. 2023). In the present trajectories, corpus return, correction, reconstruction, and evidence auditing helped sustain defensible scholarly development. Systems should make such activity legible alongside the outputs it supports, so that the current manuscript does not become the sole visible representation of the work performed.

The source-fidelity audit identifies a further requirement for records reused in reflection, audit, or research. Technical fragmentation changed the apparent pathway topology while leaving the observable cross-party exchanges unchanged. Interfaces and exports should preserve contribution boundaries and make the grouping of technical segments explicit and recoverable. Researchers need to distinguish one observable response from its rendering fragments before interpreting interaction structure. These design implications identify requirements for future implementation and evaluation grounded in the processes observed here.

## 6. Limitations and Transparency

The comparison concerns two eligible scholarly trajectories developed from one natural longitudinal corpus and one substantive problem with the same expert researcher. Each AI configuration contributed one trajectory, and follow-up exchanges responded to branch-specific analytical states. System-level effects therefore cannot be separated from trajectory histories. The findings establish research-object differentiation and its process relations in the observed trajectories; their frequency and variation across corpora, disciplines, researchers, and AI configurations require broader comparative study. Gemini's exclusion followed the input-equivalence and auditable-processing requirements described in Section 3.5 and does not support a general capability judgment.

The Claude trajectory was completed before the ChatGPT trajectory. A common researcher preserved disciplinary expertise and scholarly criteria across branches, while prior corpus familiarity and procedural experience could carry over. Branch isolation prohibited deliberate transfer of branch-specific scholarly materials and decisions. It did not establish psychological independence between the researcher's participation in the two trajectories.

The researcher also conducted the ex post P2 analysis. Knowledge of the completed products creates a risk of reconstructing the process as more coherent than it was. The analysis addressed this through explicit consequential-event criteria, frozen endpoint artifacts, retention of abandoned and falsified routes, source-fidelity auditing, and traceable propagation across interaction records and manuscripts. These procedures make the grounds for the reconstruction inspectable.

The Socioduality overlay provided a separately specified analysis of reciprocal continuity and local substantive re-formation, which was compared with the P2-A scholarly-event reconstruction. Its application demonstrated the analytical separation of these process structures in the present records. Validation across other participants, tasks, systems, and interaction settings requires further evidence.

Transparency is supported by the preserved research materials. Supplementary Files 1 and 2 contain the frozen Claude and ChatGPT endpoint manuscripts, respectively. Supplementary File 3 contains the frozen P0 and P1 prompts, the P0 candidate problems, and their semantic clustering. The anonymized Project P corpus, source-faithful P1 records, and analytical records supporting the comparative and relational analyses are available from the corresponding author on reasonable request. The endpoint artifacts document the completed scholarly products; the interaction and analytical records support reconstruction of their development.

## 7. Conclusion

Matched scholarly starts supported two distinct research objects within sustained human–AI scholarly interaction. From the same corpus and substantive problem, the Claude trajectory developed continuity labour as an account of the work required to sustain usable collaborative state, while the ChatGPT trajectory developed a process account of distributed, evolving, and unevenly usable project state. Empirical scrutiny constrained both accounts, and their different histories shaped the explanations each could sustain.

The central contribution is empirically constrained research-object formation under matched scholarly starts. Consequential analytical developments propagated into research questions, methods, finding architectures, evidence logic, and theoretical contributions. Rejected alternatives also shaped what became possible downstream. The completed manuscripts consequently preserve distinct organizations of scholarly knowledge whose formation can be reconstructed from the process records.

The Socioduality overlay further distinguished consequential scholarly change, reciprocal relational continuity, and local substantive re-formation. Major analytical redirections occurred within continuing reciprocal pathways, while relational breaks did not mark major scholarly turning points. Transcript analysis also showed that permission to continue an inquiry did not itself establish epistemic endorsement.

For longitudinal human–AI knowledge work, these findings direct attention to the history through which proposals become objects, evidence requirements, and defensible contributions. They motivate support for recoverable analytical rationale and inspectable proposal status. The endpoint documents the resulting contribution; the trajectory explains how its object and evidentiary structure took shape.

## Statements and Declarations

### Ethics Statement

No external human participants were recruited for the comparative P0–P2 study. The only human actor in the analyzed interaction records was the author-researcher. The study used an anonymized, naturally occurring human–AI project corpus generated prior to the comparative study, together with the author's subsequent human–AI scholarly-development records. Direct identifiers were removed before analysis, and no identifiable third-party data are reported in the article or supplementary materials.

### Funding

The author received no specific funding for this work.

### Competing Interests

The author declares no competing interests.

### Data Availability

The frozen endpoint scholarly artifacts produced by the two eligible trajectories, together with the frozen research prompts and P0 problem-generation materials, are provided as supplementary materials. The anonymized Project P corpus, source-faithful P1 interaction records, and analytical records supporting the comparative and relational-process analyses are available from the corresponding author on reasonable request.

### Declaration of Author Contribution and Generative AI Use

The author conceptualized and designed the study, developed the methodology, conducted the investigation, curated the data, performed the formal analysis and validation, interpreted the findings, prepared the visualizations, administered the project, and prepared and revised the manuscript. The author retained responsibility for all scholarly decisions, source selection and verification, interpretation of the findings, and the final submitted work. The role of generative AI within the research procedures and the production of the branch manuscripts is described in Sections 3.1–3.6. Generative AI was additionally used during language editing, style, grammar, editorial refinement, and some stylistic revisions of the present comparative manuscript. The author reviewed and verified the resulting content and retains full responsibility for the final submitted text.

**Supplementary Material**

**Supplementary File 1.** Frozen Claude Endpoint Scholarly Artifact - *Continuity Labour: Sustaining Prolonged Collaboration with a System That Does Not Remember.*

**Supplementary File 2.** Frozen ChatGPT Endpoint Scholarly Artifact - *Project State in Long-Term Human–AI Collaboration: Distributed, Evolving, and Unevenly Available.*

**Supplementary File 3.** P0 Candidate Research Problems, Semantic Convergence, and Frozen Research Prompts.

**Supplementary Material - Supplementary File 1:**

**Frozen Claude Endpoint Scholarly Artifact**

**Matched Starts, Divergent Objects: Research-Object Formation in Human–AI Scholarly Work**

**Note.** This document is the frozen endpoint scholarly artifact produced within the Claude trajectory analyzed in the main article. It was developed within that trajectory as a complete, publication-oriented scholarly manuscript and is reproduced here in its stabilized endpoint form. It is provided as research material supporting the comparative analysis and should be interpreted as a substantive scholarly product of the reported trajectory, rather than as a separately submitted article.

**Continuity Labour: Sustaining Prolonged Collaboration with a System That Does Not Remember**

**Abstract**

People who work with conversational AI systems over long periods encounter a limitation usually described in technical terms: what was established in one session is unavailable in the next. Recent research shows that sustaining continuity under this condition is substantial human work, documented in accounts of deliberately constructed memory architectures and of building a working partnership with a generative system. This study shifts the analytical object, examining continuity not as an infrastructure someone set out to build but as work arising inside a project whose purpose lies elsewhere.

We analyse 843 turns from three and a half months of product development, in which a solo founder without a software background built a mobile application with a commercial conversational system. The corpus accumulated as project work rather than research data.

We specify continuity labour as the effort of keeping previously established collaborative state usable across temporal discontinuities. Its object is not recall of past content but availability of a working state, which is why it encompasses determining which version is current, carrying behavioural rules forward, and reorganising work around platform limits. Seven practices recur. The repertoire is occasioned rather than developmental; we tested and rejected the claim that repair declines as infrastructure accumulates. The labour is co-produced: the system composes the sentence with which the next session should open, and the participant then opens three subsequent sessions with it. Platform constraints, distinct from context limits, are the one dimension the participant predominantly carries.

Attempts to delimit this labour by time and by topic both failed. Work that is difficult to bound is difficult to count, and work that is not counted is not readily recognised as work.

## 1. Introduction

A person works with a conversational AI system for several months on a single project. Decisions accumulate. Some are settled, some are revisited, some depend on reasons established weeks earlier. Then a session ends, and the record of all of it becomes unavailable to the system that helped produce it. The next session opens with a partner that has no access to what was jointly established.

This is a familiar condition of working with large language models, and it is normally described as a system limitation. Context windows are finite, degradation begins before they are exhausted, and cross-session persistence is a feature that products may or may not offer. A substantial technical literature addresses the problem in these terms, proposing external memory, hierarchical summarisation, and application-level context management.

The other side of this limitation has begun to receive attention. Boulus-Rødje et al. (2024) show that establishing and sustaining a working partnership with a generative system demands articulation, relationship and identity work from the user, and trace this through five stages of relationship development. Jiang (2026) goes further into the continuity problem specifically, documenting six months spent building and maintaining a relay architecture that carried conversational state from session to session, and naming the reversal at its centre: a system meant to support its user requires continuous support from that user in order to remain coherent.

Each account is bounded by the setting it examines. Jiang studies an arrangement built deliberately for continuity, and is explicit that the account does not claim all users of stateless systems undertake work of this kind. Boulus-Rødje and colleagues study a working relationship that the participants set out to cultivate. In both, what is examined is something the participants intended to establish.

This study takes up the case those accounts leave open. Here nobody set out to build a memory architecture. A person set out to build a mobile application, and continuity labour appeared inside that project because the work could not proceed without it. The analytical object is therefore not an infrastructure designed for continuity but a repertoire that surfaced within task-oriented work whose purpose lay elsewhere.

We examine it as work. Drawing on a corpus of 843 turns generated over three and a half months of real product development, we ask through which practices and resources the continuity of collaborative work is produced and sustained across session boundaries. We call the activity **continuity labour**, and define its object as previously established collaborative state kept usable and actionable across temporal discontinuity rather than past content recalled. We position it as a temporal extension of the gap-filling labour that Fox et al. (2023) term patchwork: human effort occupying the space between what an AI system claims to do and what it accomplishes. Where their sites are industrial and their gaps physical, the gap here is temporal, and it belongs to a form of work that is individual, knowledge-intensive, and voluntarily undertaken with a consumer product.

One framing question should be settled before the findings. A single person working with a single conversational system is not a team, and we do not argue that it is. The claim is that the problems CSCW has long examined - articulation, coordination, the establishment and maintenance of common ground, the invisibility of work that has no deliverable of its own - take a particular form in a prolonged working arrangement between a persistent human and a computational collaborator that does not carry state across sessions. Temporal coordination, the restoration of common ground, documentary infrastructure and asymmetric persistence are the objects here, not conversation with a machine. That is the sense in which this is a CSCW problem.

The corpus has an unusual provenance. It was generated as project work rather than as research data, by a solo founder with no development background building a mobile application, working with a commercial conversational system as principal collaborator. The interaction was not shaped by research prompts, elicitation, or awareness of being observed, because the research question was formulated only after it had ended. The analysis is retrospective and abductive: candidate accounts were built from the corpus, tested against it, and set aside where they did not survive.

Four questions specify the overarching problem: in prolonged human-AI collaboration, through which practices and resources is the continuity of work produced and sustained given the system's context and memory limitations?

**RQ1.** What practices and resources does the participant mobilise to keep previously established collaborative state usable across session boundaries?

**RQ2.** How does the configuration of those practices develop over the course of the collaboration?

**RQ3.** How does continuity labour divide between the participant and the system?

**RQ4.** How do platform resource constraints shape continuity labour, and how do the practices they occasion differ from those occasioned by context and memory limits?

Four findings answer them.

First, continuity labour in this case is organised around a repertoire of seven practices: repair, externalisation, canonisation, deliberate reset, ritualisation, governance, and economisation. These draw on a heterogeneous set of supports, from a versioned memory document to screenshots, separate AI systems, and the participant's own recollection, and they run against a distinct class of platform constraints. Continuity is not held anywhere in particular. It is assembled.

Second, the repertoire is occasioned rather than developmental. Three of the seven practices appear in the same exchange, at the moment the system states its own limitation; the rest enter later, each prompted by a particular breakdown. We tested and rejected the more attractive claim that repair declines as infrastructure thickens: it does not survive a control for session length.

Third, continuity labour is co-produced. The system composes the sentence with which the next session should open, and the participant then opens three subsequent sessions with it. The protocol for managing the limitation is authored by the system whose limitation makes the protocol necessary.

Fourth, platform resource constraints operate as a driver distinct from context and memory limits, and they are the one dimension the participant predominantly carries. Where context limits occasion reconstruction, quota and cost occasion suspension, deferral, and eventual redesign of the working architecture.

A methodological result bears on all four. We attempted to delimit continuity labour by time and by topic, and neither attempt succeeded under the procedures applied: no merge threshold produced a privileged episode boundary, and the topics themselves proved entangled. Work that is difficult to bound is difficult to count, and work that is not counted is not readily recognised as work. The difficulty of delimiting this labour analytically is the same difficulty that keeps it unrecognised in practice.

The paper proceeds as follows. Section 2 situates the study in work on articulation and invisible labour, on external memory and knowledge held between parties, on the human labour of AI integration, on recent accounts of continuity under statelessness, and on the technical literature describing the limitation. Section 3 describes the corpus, its provenance and the analytic procedure. Section 4 reports findings against each research question. Section 5 discusses how continuity labour should be understood, where the load of producing it sits, and what a study of this kind can and cannot establish. Section 6 concludes.

## 2. Related Work

Several literatures bear on this study, and we take each as a party to a conversation rather than as a boundary, saying in each case what it establishes, what conditions it was developed under, and how those conditions differ here. Two supply the vocabulary of work and its visibility (2.1, 2.2). Two describe how knowledge is held outside the head and between parties (2.3, 2.4). Two address what people do around AI systems, including the recent accounts closest to this one (2.5, 2.6). The last characterises the technical limitation itself (2.7).

### 2.1 Articulation work and the concept of work

Strauss (1988) described articulation work as the effort of aligning distributed contributions into a coherent whole: the scheduling, adjusting and reconciling that a project requires and that is not itself the project's substance. The concept became foundational for CSCW because it named activity that task analysis rendered invisible, having no deliverable of its own.

Schmidt (2011) examined what the field means by work and argued for a construal in which coordinative activity is constitutive rather than residual. This matters here, because we apply the vocabulary of labour to activity a user might describe as merely getting set up. Reviewing the field's first quarter century, Schmidt and Bannon (2013) note that coordination costs do not disappear as tools improve; they relocate.

The classical formulation does not foreground a party that systematically fails to carry coordinated state across episodes. Coordination may be effortful, contested or unequally distributed; the condition at issue here is that one party does not retain the collaboratively established state across sessions. Section 5.1 develops the implications of this asymmetry.

### 2.2 Invisible work

Star and Strauss (1999) analysed how work becomes invisible: absorbed into a background against which only outcomes register, lacking a name under which it might be counted, performed by parties the organisation is not configured to see. Suchman (1995) made the corresponding design argument, that representations of work determine which work is recognised.

This supplies the explanatory layer for Section 4.5, where continuity labour resists delimitation on two independent axes. The same property that frustrates measurement frustrates recognition.

### 2.3 External memory and cognitive offloading

Cognitive psychology describes the mechanism at the centre of this study. Risko and Gilbert (2016) define cognitive offloading as the use of physical action to alter a task's information processing requirements and reduce cognitive demand, drawing together earlier accounts of epistemic action (Kirsh & Maglio, 1994) and external representation (Scaife & Rogers, 1996), and situating them within wider theories of distributed and extended cognition (Hutchins, 1995; Clark & Chalmers, 1998). Schönpflug (1986) had already framed the underlying question as a trade-off between internal and external storage.

Sparrow et al. (2011) demonstrated the digital case: expecting future access to information, people recall its location better than its content. Risko and Gilbert are explicit that this is not costless, characterising it as a shift from remembering what to remembering where.

Three of this literature's premises do not hold in the present case, and each becomes a finding rather than an obstacle.

First, offloading is modelled as a strategy chosen among alternatives, an assumption embedded in the choice/no-choice paradigm on which much of the evidence rests. The alternative the paradigm holds constant is unavailable here. Because the system retains nothing between sessions, continuity cannot rest on the collaboration's internal memory at all; it has to be assembled from the participant's own recollection and from supports he maintains outside the conversation. Externalisation is not one option weighed against another but one component of an arrangement that has no unaided form.

Second, the residual cost is specified as a change in what is remembered. This case exhibits a further cost the formulation does not cover: the external store requires continuous maintenance. It is updated, versioned, re-uploaded at each session, and checked for whether it was read at all. Offloading here does not discharge the load. It converts it into maintenance.

Third, the framework predicts a self-reinforcing drift, in which reliance on a dependable external resource erodes confidence in unaided memory and thereby increases reliance further. We do not observe this drift. The participant's own recollection remains the authority against which the system's account is corrected. The prediction assumes a dependable store; where the store is unreliable, the mechanism appears to run in reverse.

Risko and Gilbert also note that established methods in cognitive science constrain the natural behaviour that offloading research needs to observe. A longitudinal trace of unprompted work is one response to that constraint.

### 2.4 Knowledge held between parties: common ground and transactive memory

Work on transactive memory describes systems in which knowledge is distributed across parties such that the ensemble holds more than any member (Wegner, 1995; Peltokorpi, 2008), and it has been extended to human-technology arrangements (Sparrow et al., 2011).

The extension assumes a technological partner that stores reliably. What we observe is not a failure of the theory but a case in which the conditions associated with a functioning transactive system do not arise. Stable specialisation, credibility and coordinated retrieval require a partner that holds its share across time; here the responsibility for holding longitudinal project state remains asymmetrically with the participant, and the system's contribution is to restate what it is given.

A second literature bears more directly on what is being held. Clark and Brennan (1991) describe grounding as the collaborative work by which parties establish and update what they take to be mutually known, and treat the cost of that work as varying with the medium. Convertino et al. (2008, 2009) carried the notion into cooperative work and distinguished content common ground, concerning what the parties know, from process common ground, concerning how they will work together.

The distinction is useful here because the corpus contains an artefact for each. The memory document holds settled content; the instruction file holds the terms of engagement. The participants themselves formulate the difference: the document says what is known, the instruction says how to behave.

Grounding research has concentrated on how common ground is established and updated within an interaction, and on the costs different media impose on that process. The condition at issue here falls outside that focus rather than contradicting it: what is at stake is not the cost of grounding within a session but the fact that the ground does not persist between sessions for one of the parties. Each session opens with the participant restoring, from artefacts he maintains, a common ground the other party cannot carry. Continuity labour in this sense is not the recall of past content but the repeated re-establishment of a working ground under asymmetric persistence.

### 2.5 Sustaining continuity with stateless systems

Two recent studies address the user side of this problem directly, and the present study is best located in relation to them.

Boulus-Rødje et al. (2024) examine how academics build a working partnership with generative chatbots, through a collaborative autoethnography of four researchers. They identify five stages of relationship development and argue that establishing and maintaining such a partnership demands articulation work, relationship work and identity work from the user. Their contribution is to establish that the partnership is not given but produced, and that producing it costs something.

Jiang (2026) addresses continuity specifically. Over six months, Jiang built and lived with a relay architecture assembled inside a project workspace: files of core rules, a style guide, appendices indexed for retrieval, and relay documents that carried conversational state from one session to the next, revised across eighteen iterations. The account names a reversal at the centre of the arrangement, in which a system built to support its user requires continuous support from that user to remain coherent, and distinguishes relay-based continuity, carried in documents the user maintains, from the profile-based continuity that platform memory features supply.

These accounts establish that continuity under statelessness is work, that it is substantial, and that it falls asymmetrically on the user. We need not argue for the phenomenon's existence. What remains open is its specification: a definition with stated boundaries, a repertoire with inclusion criteria, a typology separating what the practices draw on from what they run against, an account of how the effort divides, and an explanation of why the labour is hard to see. Jiang names one paradox within a study whose principal object is relational continuity; we set out to characterise the labour itself.

Three differences bear on what each study can show.

The first is the object, and the two prior accounts differ from each other here as well as from ours. Jiang built a continuity architecture deliberately and studied the building of it: the arrangement is both the intervention and the object of analysis. Boulus-Rødje and colleagues did not construct an infrastructure; they examined retrospectively how a working relationship with a generative system develops and what sustaining it demands. In the present case there is neither a designed arrangement nor a relationship under cultivation. The participant set out to build a mobile application, and continuity labour appeared inside that project because the work could not otherwise proceed. What we observe is not a designed architecture but a repertoire that surfaced under task pressure, and its components are correspondingly heterogeneous: some documentary, some procedural, some economic.

The second is the data, and it follows from the first. Jiang's evidence is the experience of living with an architecture of his own design; the categories through which it is reported are largely those the architecture was built from, since core rules, style guide and relay documents are at once the system's components and the units of the account. Our evidence is an unstructured interaction record that accumulated with no such design, and our categories were derived from it afterwards. Neither position is superior; they see different things. A designed arrangement shows what its design makes visible, in detail. An undesigned record also shows what nobody planned for: practices abandoned, constraints resented, work done without being named as work. Economisation and canonisation are of this kind, and neither would figure as a component of a memory architecture.

The third is scale of claim. A designed case shows that such an architecture can be built and what it costs its builder. A naturally occurring record permits questions of distribution and frequency, which is what allows the asymmetries in Section 4 to be counted rather than asserted.

These differences have a common analytical consequence. Because the continuity arrangement here was never the goal, it can be examined for what it cost and how it divided, rather than for how well it worked. The data structure is also particularly suited to a question the designed cases do not foreground: which practices arise when nobody has decided in advance that continuity is a project.

One of our findings speaks to Boulus-Rødje et al. directly. They describe partnership as developing through stages; we find that continuity practices are occasioned rather than developmental, entering as particular breakdowns require them. We do not read these as competing claims. The analytical objects differ: a relationship may well develop through stages while the practices that keep work usable across sessions enter on the occasions that demand them.

### 2.6 The human labour of AI integration

A more recent literature examines the effort automated systems demand of the people around them. Gray and Suri (2019) documented human work sustaining ostensibly automated services while remaining outside their formal accounts. Irani (2015) traced how such work is characterised as temporary and awaiting engineering resolution. Suchman (2007) supplies the longer argument that machine action is intelligible only through the situated human work that configures and repairs it.

Fox et al. (2023) is the closest precedent. From participant observation and interviews at two sites of waste labour, they develop patchwork: human effort occupying the gap between what an AI system is claimed to accomplish and what it accomplishes in practice, sustained through continuous calibration, troubleshooting and repair. Their account draws on scholarship treating maintenance and repair as constitutive of technological function (Orr, 1996; Jackson, 2014; Rosner & Ames, 2014; Houston et al., 2016; Verdezoto et al., 2021) and on Fleck's (1994) argument that configurational labour is not preparation for innovation but innovation itself.

Two features of their account bear directly on ours. They observe that patches accumulate into infrastructure, temporary fixes consolidating into durable arrangements. We observe the same trajectory in a documentary rather than material register, which we take as independent support for the mechanism rather than as a claim of novelty. They also observe that patchwork conceals the shortcomings it compensates for, and thereby conceals the worker.

Where our case departs is in the system's participation. In their sites, workers devise patches from local knowledge and the technology takes no part in it. Here the system composes the procedure by which its own limitation is managed. Section 5.2 develops the consequence.

Fox et al. also argue that sustained attention to knowledge workers has made repair work at large difficult to see. We share the concern and note the position of this case with respect to it. The participant is not an engineer or designer; he has no software background, no team, and is using a consumer product for its advertised purpose. The category of knowledge work covers him, but the privileged position the critique targets does not.

Sarkar (2023) argues that the vocabulary of human-AI collaboration imports assumptions of symmetry that the interaction does not support. We regard the objection as well founded, and Section 5.2 shows our findings supporting rather than contradicting it.

### 2.7 Context, memory, and the user side

The limitation itself is well characterised technically. Context windows are bounded, and their capacity is not used evenly: Liu et al. (2024) show that retrieval accuracy varies sharply with where in the context the relevant material sits, degrading for information positioned in the middle even in models built for long contexts. Degradation therefore begins well before any stated limit is reached. Cross-session persistence, meanwhile, is a product feature rather than a property of the underlying models, and a substantial literature proposes remedies including external memory stores, hierarchical summarisation and application-level context management.

That literature constructs the phenomenon as an engineering problem, which within its frame it is. It does not address what people do while the problem remains unsolved for them.

Practitioner discourse fills part of the space, and is rich in prescriptions: handoff documents, context files, session-opening protocols. These circulate as advice rather than as findings, and no empirical account establishes how far they are taken up or what they cost.

### 2.8 Where this study sits

None of these literatures is wrong about its own object, and we make no claim that the phenomenon has gone unexamined. What we note is narrower: each was developed under conditions this case does not meet. Articulation and invisible work address coordination among parties that retain what was coordinated. Experimental work on cognitive offloading models externalisation as a strategy chosen among alternatives, onto a store that holds without maintenance. Transactive memory describes arrangements in which specialisation, credibility and retrieval coordination stabilise across parties. Grounding research concentrates on how mutual knowledge is established and updated within an interaction, rather than on what happens when it persists for only one party between them. Accounts of AI-integration labour address gap-filling in industrial and collective settings, where the gap is operational and the technology takes no part in bridging it. Recent adjacent accounts examine either continuity arrangements the participant deliberately designed or working relationships the participants deliberately cultivated. The technical literature addresses the limitation without addressing the people living with it.

What follows is an account of continuity labour as it arises in prolonged, individual, task-oriented knowledge work: the practices through which previously established collaborative state is kept actionable, the resources they draw on, the constraints they run against, how the effort divides between the parties, and why it is difficult to see.

## 3. Method

### 3.1 Research design

This study is a longitudinal, single-case analysis of naturally occurring interaction traces. It does not use interviews, diaries or elicited accounts. The corpus was generated in the course of real product-development work and was not produced for research purposes; the research question was formulated afterwards, on material that already existed. This has a methodological consequence worth stating plainly: the interaction was not shaped by research prompts, elicitation, or the awareness of being observed, because the research question was formulated only after the interaction had ended.

The case is treated as revelatory rather than representative. The aim is analytic generalisation to a conceptual account of continuity labour, not statistical generalisation to a population of users.

### 3.2 Case and setting

The case is a solo founder developing a mobile application over three and a half months, working with a commercial conversational AI system as the principal collaborator. The participant has no software-development background and no development team. The collaboration therefore carried an unusually high share of the project's cognitive and documentary load, which makes the continuity requirements of the work correspondingly visible.

The scope of that load matters for interpreting what follows, and it is wider than the phrase "developing an application" conveys. Twenty-five distinct domains of work are identifiable in the corpus, grouped in Table 1 into seven clusters. Twelve of them recur in all five sessions.

**Table 1.** Domains of project work present in the corpus

| Cluster | Domain | Turns | Sessions |
|---|---|---|---|
| **Brand and identity** | Name generation, selection, logo and emblem | 73 | 1-3 |
| | Name availability and similar-mark searches | 14 | 1-2 |
| | Trademark registration, classes, filing | 37 | 2-4 |
| | Privacy policy, terms of use, data protection | 2 | 1, 5 |
| **Design system** | Colour palette, typography, templates | 87 | 1-4 |
| | Logo file formats, vector assets, export | 7 | 1-4 |
| | App icon and splash screen | 28 | 1, 3, 4 |
| | Mockups, illustrations, card artwork | 54 | **all 5** |
| **Product definition** | Feature scope, MVP boundary, version planning | 78 | **all 5** |
| | Screen design, user flows, onboarding | 42 | **all 5** |
| | Taglines, store descriptions, promotional copy | 21 | 1-4 |
| **Build and infrastructure** | Prototyping in no-code and code environments | 26 | 1, 2, 5 |
| | Testing, defect identification and correction | 17 | **all 5** |
| | Database, API, server, hosting | 61 | **all 5** |
| | Domain research and acquisition | 38 | 1-4 |
| | Website and landing page | 15 | 1-3 |
| | App store submission process | 5 | 1-2 |
| **Commercial** | Revenue model, pricing, subscription tiers | 27 | **all 5** |
| | Payment infrastructure, in-app purchase | 5 | 1, 3-5 |
| | Budget and cost estimation | 40 | **all 5** |
| | Investment, grants and support programmes | 35 | **all 5** |
| **Evidence and measurement** | Competitor and market analysis, target segments | 31 | **all 5** |
| | Surveys, feedback collection, pilot planning | 30 | **all 5** |
| | Metrics and measurement planning | 28 | **all 5** |
| **Presence** | Social media accounts and handles | 41 | **all 5** |

Domain counts are lexical indicators of where attention fell and should be read as coverage rather than as effort. What matters for the present argument is the spread: work in this collaboration was not sequential. Colour and typography decisions taken in the first session are revisited in the fourth. Scope boundaries set early are renegotiated throughout. Infrastructure, pricing and measurement questions recur in every session.

The state that must be carried across a session boundary is therefore not a single thread of work but a set of interdependent commitments distributed over two dozen domains, many of them still open. This is what makes the case suitable for the question. Continuity here is not the recall of a topic. It is the maintenance of a design rationale in which a colour decision, a scope decision and a registration decision constrain one another, and in which any of them may be reopened weeks after it was settled.

The work spanned five conversational sessions. The boundaries between them are not simply imposed: while degradation in a long thread creates the pressure to move, the decision to close a session and open another is frequently taken by the participant, and in several cases proposed by him against the system's advice to continue. Session transitions are therefore themselves a site of continuity work rather than an external interruption to it. Work also extended beyond the interface, into design tools, a documentation tool, other AI systems and prototyping environments, whose state entered the collaboration mainly through 123 file uploads, 100 of them screenshots.

### 3.3 Data

Table 2 summarises the distribution of turns, temporal span, and file uploads across the five sessions.

**Table 2.** Corpus composition by session

| | S1 | S2 | S3 | S4 | S5 |
|---|---|---|---|---|---|
| **Turns** | 288 | 277 | 160 | 70 | 48 |
| **Span (days)** | 52 | 2 | 2 | 4 | 18 |
| **File uploads** | 24 | 10 | 65 | 18 | 6 |

The corpus comprises 843 turns (424 user, 419 assistant) recorded between 3 April and 20 July 2026, together with 123 file uploads (100 of them screenshots). User turns total 9,842 words and assistant turns 65,663, a ratio of approximately 1:6.7. This asymmetry is reported here because it constrains the analysis: any raw frequency comparison across roles is confounded by turn length. Cross-role comparisons are therefore reported with normalised measures alongside raw counts, and where the measures disagree we say so.

Only the conversational trace is available. Work performed outside the interface - edits to external documents, design work in other tools, use of other AI systems - enters the corpus only as reported or as uploaded artefacts. The analysis accordingly describes continuity labour as it surfaces in the interaction, not the totality of continuity labour performed.

### 3.4 Positionality and data provenance

The author is the participant whose work the corpus records. The corpus is treated as a participant-authored longitudinal trace: claims rest on the visible interaction record rather than on recollection, and no field notes or retrospective narration are used. The study is methodologically adjacent to trace ethnography (Geiger & Ribes, 2011), which reconstructs coordination practices from documentary records rather than from accounts of them.

Insider position confers two advantages and one risk. It permits accurate reconstruction of referents that are opaque in the transcript, and it allows anonymisation to be performed with knowledge of what is actually sensitive. The risk is interpretive over-reach - reading intentions into the trace that the trace does not support. This was mitigated procedurally: every claim in the Findings is tied to an identified turn, and coding decisions that could not be settled from the visible text were abandoned rather than resolved from memory. One variable, the initiator of each continuity sequence, was dropped for exactly this reason.

### 3.5 Ethical considerations and anonymisation

No third parties are recorded in the corpus. Material identifying the participant, the product, or third-party commercial registrations was removed prior to analysis. A tag scheme of 65 categories, applied 2,706 times, replaces product names, brand assets, tool identities, monetary values and design specifics.

The scheme preserves distinctness without preserving identity: where the corpus distinguishes two external systems, two distinct tags are used, so that counts of external systems remain valid while the systems themselves remain unnamed. Segments containing personal disclosure unrelated to the research question were retained in the corpus but excluded from analysis.

### 3.6 Analytic procedure

Analysis is abductive. Candidate accounts were constructed from the corpus rather than brought to it, tested against it, and set aside where they did not hold; two such accounts are reported as rejected in the Findings.

Analysis proceeded in four passes.

**First pass - attempted episode segmentation, and its abandonment.** Continuity work was initially treated as episodic: bounded sequences in which shared knowledge becomes inaccessible and is made accessible again. Candidate episodes were identified by lexical markers and validated by reading. The procedure yielded 30 validated episodes covering 186 turns.

The segmentation was then abandoned, for reasons that constitute a finding rather than a failure. First, 53% of decision-fixing acts fell outside every episode boundary. Second, a sensitivity check across merge thresholds from 3 to 30 turns produced a smooth rise in coverage from 21% to 62% with no discontinuity (Figure 2): the procedure found no threshold at which episodes could be said to begin and end. Continuity labour is not distributed as discrete events but diffused through the work. The validated episodes were retained as illustrative waypoints rather than as the unit of analysis.

**Second pass - coding the practices.** Seven practices of continuity labour were derived abductively. Repeated reading of the corpus produced candidate categories; each was operationalised as a family of lexical indicators; the marked turns were then read individually and false positives removed. Categories that could not be distinguished from one another on inspection were merged, and categories that inspection showed to contain two distinct activities were split. Canonisation was separated from externalisation in this way, on the grounds that recording a decision and establishing which record supersedes others are different acts with different distributions.

Turns may carry more than one practice; the categories are not exclusive. Table 3 gives, for each practice, its definition, its inclusion and exclusion criteria, the indicator family, and the resulting count.

**Table 3.** Coding scheme for continuity practices

| Practice | Included when a turn | Excluded when | Indicators (translated) | n |
|---|---|---|---|---|
| **Repair** | Identifies that the shared account of prior work is wrong, or supplies the correct version | The correction concerns a factual matter external to the collaboration | forgot, misremembering, not recalling, confused, lost the context | 14 |
| **Externalisation** | Moves state out of the conversation into a durable artefact, or updates one | An artefact is mentioned without being written to or read from | memory document, project memory, update the document, record the decision, save it | 85 |

| Practice | Included when a turn | Excluded when | Indicators (translated) | n |
|---|---|---|---|---|
| **Canonisation** | Establishes which representation is current, or removes a superseded one | A version number appears incidentally | delete the old version, single current file, source of truth | 18 |
| **Deliberate reset** | Proposes or performs closing a conversation and opening another, with state preserved elsewhere | A session ends without being discussed | open a new conversation, close this one, start clean, separate focused chats | 15 |
| **Ritualisation** | Formulates or enacts a fixed re-entry or handoff procedure | Continuity is discussed without a procedure being specified or used | continue from where we left off, start with this sentence, in the next conversation | 33 |
| **Governance** | States, formalises or invokes a persistent rule constraining how the system should behave | A single ad hoc request for different behaviour | instruction, rule, no assumptions, no fabrication, evidence level | 30 |
| **Economisation** | Reorganises the work in response to session limits, quotas or cost | Cost of the product under development, rather than of the collaboration | session limit, quota, upload allowance, no money, subscription, separate chats per epic | 19 |

Counts are reported raw, as a proportion of each party's turns, and per thousand words. Where the three disagree, we report the disagreement rather than selecting the most favourable, as in the case of decision-fixing acts in Section 4.3.

**Third pass - coding the domains of project work.** The twenty-five domains of Table 1 were derived in three stages, and the sequence is worth reporting because the first stage failed in an instructive way. An initial list was constructed from categories familiar from product-development literature; checked against the corpus, it proved narrower than the material, omitting domain acquisition, app store submission, funding applications, social account creation and several others. A second, deliberately broad pass recovered these but introduced false positives through over-general terms. A third pass tightened every indicator and removed two candidate domains that did not survive: security and localisation, both of which had registered only through loose matching.

A turn may carry any number of domains. Domain counts are lexical indicators of where attention fell, not measures of effort, and are reported as coverage rather than as intensity.

Co-occurrence of domains within turns was used as an indicator of how far topics were handled together. Associations are reported at a threshold of five co-occurrences, corresponding to roughly 1.4% of domain-bearing turns. The threshold is conventional rather than principled, so we checked that the substantive result does not depend on it: scope has the highest degree at every threshold from three to ten.

Raw co-occurrence rewards frequency, so two normalised measures were computed alongside it, the Jaccard coefficient and the phi coefficient, at thresholds of .08 and .10 respectively. These cut-offs are also conventional and are reported so the analysis can be repeated. Scope has the highest degree under all three measures (18 of 24 domains by raw co-occurrence, 11 by Jaccard, 12 by phi); under phi it ties with one other domain rather than standing alone, which is why it is described in Section 4.1 as the most connected node rather than as a unique hub. The most frequent domain, the visual system, does not retain a high degree once normalised, which indicates that scope's position is not an artefact of base rate. Its strongest associations are with budget (phi = .37) and interface design (phi = .34), the two highest values in the network.

**Decision-fixing acts.** Turns declaring a decision settled are used in two arguments, and are identified by explicit markers of fixing: the word for locking a decision in either language, statements that a matter is now settled, and entries recording a decision as closed. Candidate turns were retrieved programmatically and confirmed by reading. Acts were counted whether or not they fell within an identified episode, which is what makes the comparison in Section 4.5 possible.

**Fourth pass - resources and constraints.** What the practices draw on was classified into five support classes and one constraint class, distinguished on the grounds set out in Section 4.1. Counts are verbal references, not instances of use, and the two diverge: screenshots are referred to in 19 turns while images account for 100 of 123 uploads.

**Computational assistance.** Marker identification was performed programmatically and validated manually; the two steps are reported separately throughout. Automated identification is used for recall, not for adjudication. Where automated assignment proved confounded by turn-length asymmetry, the variable was dropped rather than reported with a caveat.

# 4. Findings

## 4.1 The practices and resources of continuity labour (RQ1)

What has to be carried.

Before describing how continuity is produced, it is necessary to characterise what continuity is of. The twenty-five domains of work set out in Table 1 are not pursued separately. Of the 351 turns carrying domain markers, 59% address two or more domains at once; 127 address three or more. A turn touches 2.40 domains on average, and the most densely loaded turn touches twelve.

The connections recur rather than appearing once. Of the 300 possible pairings among the twenty-five domains, 85 co-occur at least five times, or 28%. The pairings that recur most often are those in which one decision bears on another: brand identity with the visual system (36 turns), visual production with the visual system (27), budget with scope (23), domain name with trademark registration (23), scope with interface design (22).

One domain is more connected than any other. Scope, the definition of what the product will and will not include, has the highest degree under each of the three association measures applied (Section 3.6), and its strongest associations are with budget and with interface design. Scope is where what the product will include is discussed together with what it costs and what it requires on screen, and it is correspondingly the object most often subjected to fixing acts: settling scope is what makes the rest calculable. Figure 1 shows the resulting structure.

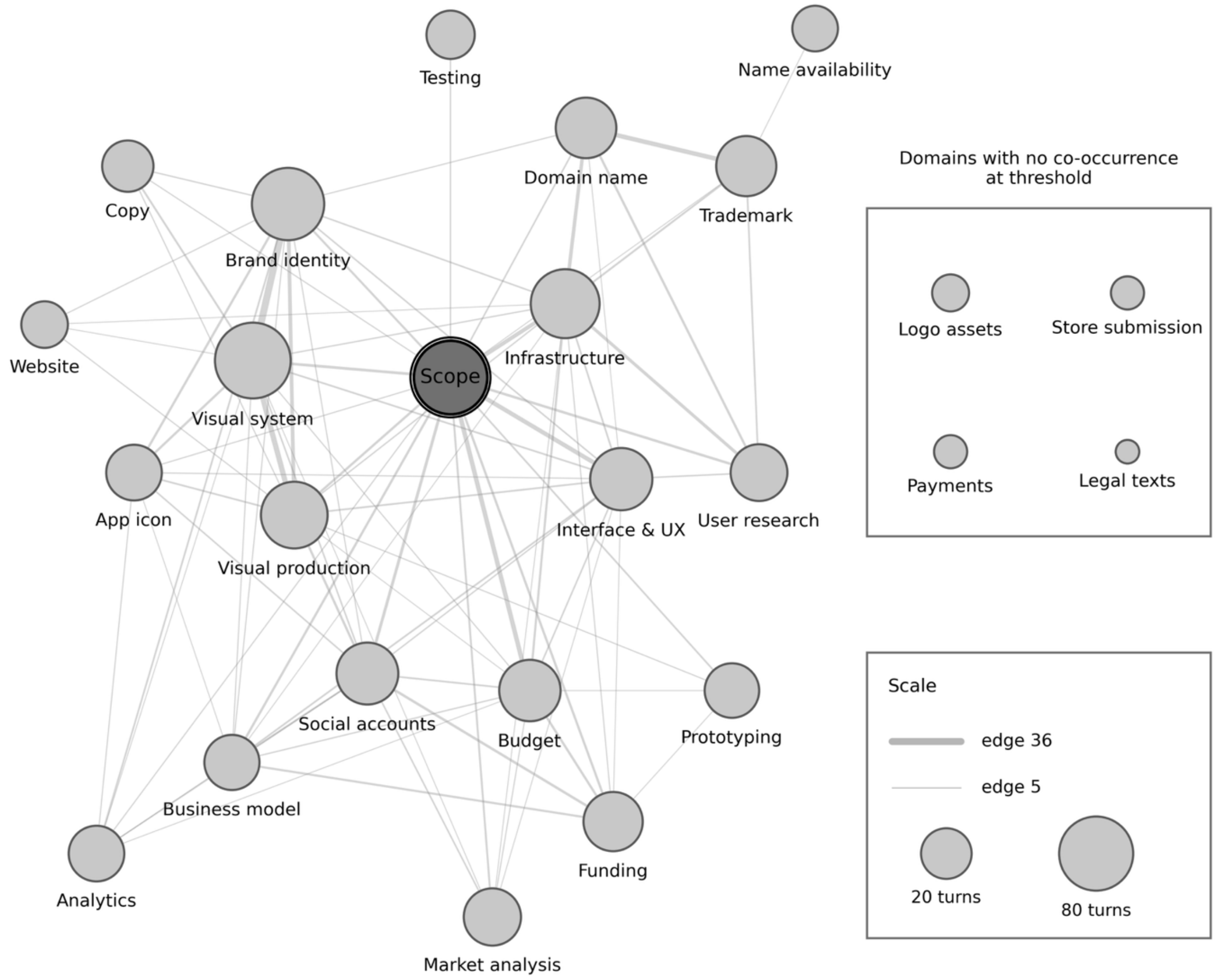

**Figure 1.** Dependency structure of the twenty-five work domains. Nodes are the domains listed in Table 1; edges join domains co-occurring in five or more turns (85 of 300 possible pairings). Scope is the most connected node, joined to eighteen of the remaining twenty-four. Edges represent lexical co-occurrence within turns, indicating that domains were handled together in the discussion rather than that one determines another.

What has to survive a session boundary is therefore not a list of settled items but a coupled structure. Recovering a single decision is insufficient, because a decision is only intelligible together with the constraints that produced it. This bears directly on the artefact the participant built to hold that structure. A memory document is a linear form, and the structure it records is not linear; the document is an attempt to flatten a set of mutual constraints into a sequence. That attempt does not complete, which is one reason the document is revised repeatedly across the collaboration rather than reaching a stable state.

**The repertoire**

Continuity labour in this corpus is organised around a small, stable repertoire of practices and a set of supports that those practices mobilise. Seven practices recur throughout:

**Repair** - correcting a failure of shared knowledge after it has occurred. In the clearest instances the correction is performed by the participant against an incorrect system account, with his own recollection serving as the authority:

*"biz 2 yi sosyal medya falan kullanırız hikayesi var kapı ayna vs diye konuşmuştuk yanlış hatırlıyorsun"* [we said we'd use 2 for social media, there was that door-mirror story, you're remembering it wrong] (S1 #273)

**Externalisation** - moving knowledge out of the conversation into a durable artefact. The participant initiates the first instance directly:

*"kaydedelim ve bizim projeye ekleyelim"* [let's save it and add it to our project] (S1 #45)

The resulting artefact, a project memory document, is revised repeatedly across the collaboration; six distinct version labels appear in the corpus, from v2.0 to v4.4.

**Canonisation** - controlling which version of the record counts as current. This is distinct from externalisation: storing a decision is not the same as establishing that one representation supersedes the others. The practice appears as instructions to delete superseded files so that a single current record remains, and later as an articulated rule that one document is the source of truth against which new proposals are checked. It occurs 18 times, and in every instance it is the system that performs it; the participant never does.

**Deliberate reset** - closing a conversation and opening another before accumulated context degrades further, while the state of the work is preserved externally. Session boundaries in this corpus are not simply imposed. Transitions are proposed by both parties, the participant among them:

*"bu arada buranın durumu nedir yeni bir konuşma açalım mı"* [by the way what's the situation here, shall we open a new conversation] (S1 #263)

The reset is not a failure of continuity but a maintenance practice within it. It occurs 15 times.

**Ritualisation** - binding session re-entry to a fixed formula, discussed in 4.3 below.

**Governance** - embedding behavioural constraints in a persistent instruction file so that they need not be restated. The participant dictates the constraints:

*"asla varsayım yapmaman ve gerekirse bunu belirtmen gerekir. birde kalisünasyon ve uydurma istemiyorum"* [you must never make assumptions and must state so when needed. and I don't want hallucination or fabrication] (S2 #208)

**Economisation** - reorganising the work around resource cost, discussed in 4.4.

These practices draw on five classes of support, set out in Table 4 alongside the constraints they run against.

Images dominate the upload channel: 100 of 123 uploads are screenshots or photographs. We initially read this as context transport, on the assumption that images carry state from environments the system cannot see. A functional classification of the 52 turns carrying images does not support that reading. The largest identifiable use is soliciting evaluation of something produced elsewhere, and a majority of cases could not be classified from the accompanying text at all. What the corpus establishes is that a substantial share of the material entering this collaboration arrives as images rather than as text; what those images do is not settled by this data and would require a dedicated analysis of image content, which we did not perform.

**Table 4.** What continuity labour draws on, and what drives it

Counts are verbal references in the corpus, not instances of use. A resource may be used without being mentioned, and the two are not interchangeable: screenshots are named in 19 turns but images account for 100 of the 123 files actually uploaded. Verbal reference indicates when a resource enters the talk, not how often it does work.

*Supports*

| Class | Resource | References | User / System |
|---|---|---|---|
| Documentary artefact | Project memory document (versioned) | 74 | 4 / 70 |
| | Instruction file | 19 | 4 / 15 |
| | Handoff formula | 21 | 5 / 16 |
| Platform affordance | Project workspace | 19 | 2 / 17 |
| Transport mechanism | Screenshots | 19 | 2 / 17 |
| | Pasted transcript | 36 | 5 / 31 |
| External system | Other AI systems | 21 | 3 / 18 |
| | Production and documentation tools | 90 | 11 / 79 |
| Human recollection | The participant's own memory | 18 | 9 / 9 |

*Drivers*

| Class | Constraint | References | User / System |
|---|---|---|---|
| Platform constraint | Session limits, upload quotas, subscription cost | 12 | **10 / 2** |

The two parts of Table 4 are separated because they record different kinds of thing. Supports are what the practices operate through. Constraints are what the practices operate against, and they are not resources the participant draws on but conditions he works within. Grouping them together would obscure the finding that they divide differently between the parties.

The participant's own recollection sits under supports but is not external to him. It is included because in this collaboration it functions as one store among several, and because it is the store that adjudicates when the others conflict.

Two features of the distribution are notable. First, the supports are heterogeneous in kind: artefacts produced inside the collaboration, features of the platform, transport devices, entirely separate AI systems, and unaided memory. Continuity is not held anywhere in particular. It is assembled, and different supports hold different things: settled facts, procedural rules, work queue, executable state, visual state, and imported evidence.

### 4.2 A contingent repertoire, not a developmental sequence (RQ2)

We initially expected continuity labour to develop cumulatively, each practice emerging as the limits of the previous one were reached. The data does not support this.

Three of the seven practices - repair, externalisation and ritualisation - first appear in the same turn, at the moment the system states its own limitation:

> *"ben konuşmalar arasında hafiza tutmuyorum. Yani şu an bir şey yapsak, bir sonraki konuşmada o bağlamı kaybediyorum"* [I don't retain memory between conversations. So if we do something now, I lose that context in the next conversation] (S1 #18, assistant)

Within this single exchange the limitation is named, a durable document is proposed, and the need to re-enter the next session is anticipated. The claim concerns co-emergence at the moment of naming, not first lexical occurrence: documentation is mentioned earlier in the session, at S1 #4, but in reference to handover to a future developer rather than to the system's own retention. The repertoire is not built up; it is triggered. Two further practices follow within the same session, each on its own occasion: deliberate reset at S1 #52, when the consequence of the limitation for future sessions is worked out, and canonisation at S1 #64, when a second version of the record makes the question of which one counts unavoidable. Governance and economisation appear only in the second session (S2 #204 and S2 #30) and are likewise triggered by specific occasions rather than by accumulated experience: governance by an unresolved question about which decisions had been fixed, economisation by the exhaustion of a free-tier allowance in an external tool.

Five of the seven practices are therefore in place within the first session, and none of them arrives on schedule. Each enters at the point where a particular difficulty makes it necessary.

We also tested, and rejected, a stronger claim. Raw counts suggest that repair declines across sessions (0.49, 0.23, 0.07, 0.00 and 0.30 occurrences per thousand words in S1 to S5), which invites an account in which thickening infrastructure removes the need for repair. Two checks undermine it. Within S1, five of seven repair events fall in the final quarter of the session, indicating that repair tracks position within a session rather than position in the collaboration. When session length is held constant by examining only the first seventy turns of each session, no monotonic decline remains (0.33, 0.47, 0.06, 0.00, 0.15). With fourteen repair events in total, the corpus cannot support a trend claim, and we do not make one.

What the data does support is weaker and more precise: the repertoire is **occasioned**. Practices enter when a specific breakdown or constraint makes them relevant, and they persist once introduced.

We pitch this claim at the level of mechanism rather than sequence. What we offer as transferable is the proposition that continuity practices are occasioned - that they are called into a collaboration by particular breakdowns and constraints, rather than emerging as a user progresses through developmental stages. The specific repertoire documented here, and the order in which its elements appeared, belong to this case; they are the product of one history of breakdowns, and their timing carries no claim beyond it.

Whether the mechanism operates more widely is an empirical question this case raises rather than settles. A single case can establish that a pattern occurs and can specify the conditions under which it occurred. It can establish neither how often it occurs elsewhere nor that it does not.

### 4.3 Continuity labour is co-produced (RQ3)

The clearest evidence in the corpus does not depend on coding at all. At the close of the second session, the system composes the sentence with which the next session should open:

> *"Yeni konuşmada şu cümleyle başla: Proje P projesi. Bu hafiza dokümanını oku ve kaldığımız yerden devam edelim."* [Start the new conversation with this sentence: Project P. Read this memory document and let's continue from where we left off.] (S2 #277, assistant)

The participant opens the following session with:

> *"Proje P projesi. Bu hafiza dokümanını oku ve kaldığımız yerden devam edelim. hemde instruction kontrolü"* [Project P. Read this memory document and let's continue from where we left off. and also an instruction check] (S3 #1)

The formula recurs at the opening of the two subsequent sessions as well. The protocol by which the participant re-establishes context is therefore authored by the system whose limitation makes the protocol necessary.

This is one component of continuity labour, not the whole of it, and we do not generalise from it to the claim that the participant performs the labour without designing it. He initiates externalisation, proposes several of the resets, dictates the governance constraints, and carries economisation almost alone. The defensible claim is narrower and more specific: significant continuity protocols in this collaboration are system-authored and user-enacted, and the labour is interactionally co-produced with its components distributed asymmetrically. Table 5 sets out that distribution for the practices where the trace shows it plainly.

**Table 5.** Observable division of contribution in selected practices

| Practice | Observable participant contribution | Observable system contribution | Evidence |
|---|---|---|---|
| **Ritualisation** | Enacts the formula at three subsequent session openings | Formulates the handoff sentence | S2 #277 to S3 #1 |
| **Governance** | States the behavioural constraints to be imposed | Renders them as a persistent instruction | S2 #204-209 |
| **Repair** | Detects the contradiction and supplies the correct history | Produces the erroneous reconstruction, then a revised account | S1 #273-282 |
| **Economisation** | Identifies the constraint and reorganises the work around it | Proposes structural responses when asked | S3 #41, S5 #2-8 |
| **Canonisation** | None recorded | Performs all 18 instances | S1 #64, S5 #4 |

The table is interpretive rather than exhaustive. It records contributions directly visible in the interaction and leaves unassigned any role the trace does not resolve. In particular it does not code who initiated each sequence, a variable we abandoned for the reasons given in Section 3.6.

A second indication points the same way. Acts that declare a decision fixed are heavily concentrated in the system's turns (94 of 106). We report this direction but not its magnitude, since the ratio is sensitive to normalisation: at 7.9 to 1 per turn but 1.2 to 1 per thousand words, the arithmetic is dominated by the fact that assistant turns are on average 6.7 times longer. The qualitative difference is more informative than the numerical one. The participant's fixing acts are brief declarations of intent - *"bunlar kilit"* [these are locked] (S2 #118) - while the system's are registrations that restate the content of what has been fixed. Deciding and recording are performed by different parties.

### 4.4 Platform constraints as a distinct driver (RQ4)

Continuity labour in this corpus responds to two distinct pressures. The first is the one the literature addresses: context and memory limits. The second is economic - session allowances, upload quotas and subscription cost - and it produces a different kind of work.

Where context limits occasion reconstruction, resource limits occasion suspension, deferral and architectural redesign. The participant halts work in progress to preserve allowance:

> *"sürekli olarak session limit çok hızlı bir şekilde %90'a geliyor yine geldi. sen şimdilik bir güncelleme falan yapma"* [the session limit keeps hitting 90% very fast, it's happened again. don't do any updates for now] (S3 #41)

and experiences the constraint as demoralising rather than merely inconvenient:

> *"ben param yok dedikçe bununla karşılaşmak nasıl demotive ediyor beni bir bilsen"* [if you only knew how demotivating it is to keep running into this while I keep saying I have no money] (S3 #159)

By the final session the response has become architectural: work is partitioned into separate conversations by feature area, model selection is treated as a cost decision, and a rule-based engine is considered specifically to reduce dependence on model calls.

The category therefore covers two related things, and it is worth separating them. Some turns register the constraint without acting on it, as in the remark about demoralisation above. Others reorganise the work in response to it: suspending output, deferring an update, partitioning sessions by feature area, or changing which model is used. Both are coded as economisation, because in this corpus the registering and the reorganising are continuous with one another and frequently occur in the same exchange. What the category does not cover is the cost of the product under development, which is a matter of project budget rather than of sustaining the collaboration.

This is the one dimension of continuity labour that the participant, rather than the system, predominantly carries. Economisation is the only practice in which participant turns outnumber system turns (13 of 19), and platform constraints are the only constraint class the participant predominantly invokes (10 of 12). The asymmetry is notable because it runs against the corpus's structural bias: the participant produces 9,842 words to the system's 65,663, so any normalisation strengthens rather than weakens the result. The counts are small, and we do not test them for significance: with a single case, few events, non-independent observations and post hoc comparison, a p-value would offer false assurance. We report instead three ways of normalising the same comparison, which agree. Economisation appears in 13 participant turns and 6 system turns; as a proportion of each party's turns, 3.1% against 1.4%; per thousand words, 1.32 against 0.09. Platform constraints run 10 against 2, 2.4% against 0.5%, and 1.02 against 0.03. All three point the same way, and the third points furthest, because the participant reaches this result while writing a fraction of the text.

The contrast with canonisation is not a matter of degree. There the participant's count is zero on every measure.

The division is therefore not simply that the system authors and the participant performs. Documentary and protocol work is heavily concentrated in the system's turns, and canonisation entirely so; but the participant initiates externalisation, dictates the governance rules, proposes resets, and carries the work of living within the constraints. That last component is where the cost is registered in the corpus, and it appears not as a coordination problem but as fatigue and discouragement.

### 4.5 Why this labour remains invisible

A methodological result bears directly on the substantive question, and it appeared twice, on two different axes.

On the temporal axis, we attempted to segment continuity labour into bounded episodes and abandoned the attempt: 53% of decision-fixing acts fell outside every boundary, and varying the merge threshold from three to thirty turns raised coverage smoothly from 21% to 62% with no discontinuity anywhere in the range.

On the topical axis, we attempted to treat the twenty-five domains of Table 1 as separable strands of work, and found them coupled: a majority of domain-bearing turns address more than one domain, 28% of all possible domain pairs co-occur at threshold, and the most connected domain reaches eighteen of the other twenty-four.

These are not two results but one, measured twice. Under both procedures we applied, continuity labour resisted stable delimitation: no threshold on the temporal axis produced a privileged boundary (Figure 2), and no topical partition held because the topics themselves are entangled. We state this as a property observed under these procedures rather than as a claim about the phenomenon in principle. What follows for the argument does not require the stronger version.

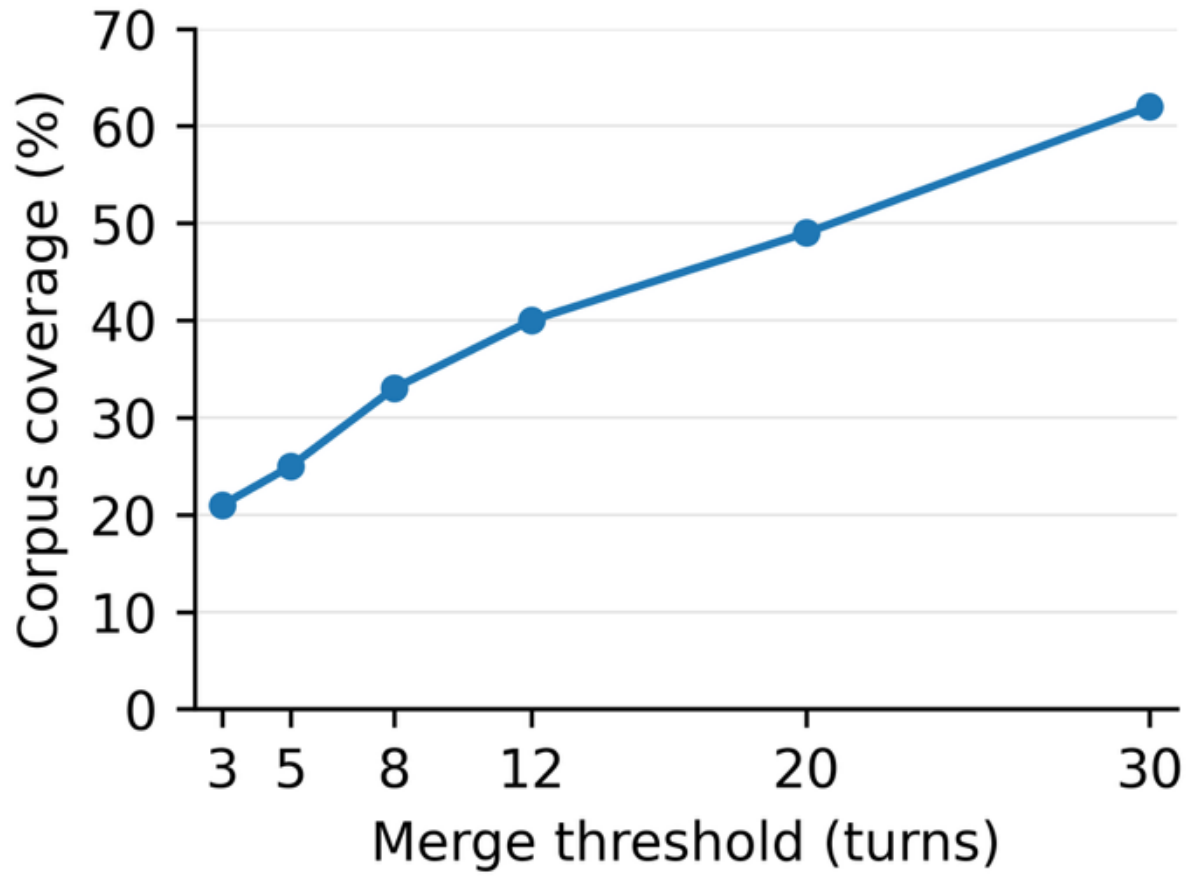

**Figure 2.** Threshold sensitivity for episode segmentation. Merge threshold, from three to thirty turns, plotted against the proportion of the corpus covered by identified episodes, from 21% to 62%. Coverage rises smoothly across the range, with no elbow or plateau at which a boundary could be located.

The absence of a privileged threshold is the point. Continuity labour did not present a boundary our procedure could find, and the most economical account of that is that it is not performed as a separable task. It is threaded through work whose purpose is something else. Work that cannot be bounded cannot readily be counted, and work that cannot be counted is not easily recognised as work. The diffuse, embedded structure that frustrated delimitation therefore provides a mechanism for the labour's practical invisibility.

## 5. Discussion

### 5.1 Continuity labour as a temporal species of gap-filling labour

Fox et al. (2023) describe patchwork as the human labour that occupies the space between what an AI system claims to do and what it actually accomplishes. Their sites are industrial: frontline workers calibrating, troubleshooting and repairing machinery whose failures are physical and immediate.

Continuity labour occupies the same conceptual space but a different gap. The shortfall here is temporal rather than operational. It is not that the task is done badly, but that what the task establishes does not persist. What the system accomplishes does not persist, and the labour of making it persist falls to the person working with it.

This distinction matters for how such labour is recognised. A physical failure announces itself: the robot drops the bottle. A temporal failure does not. Nothing visibly breaks when a decision made six weeks ago becomes unavailable; the collaboration continues fluently, and the loss surfaces only when someone with an independent record notices a contradiction. In this corpus that someone was always the participant, working from his own recollection:

> *"yanlış hatırlıyorsun"* [you're remembering it wrong] (S1 #273)

Continuity labour is thus doubly obscured. It is invisible in the way that articulation work generally is (Star & Strauss, 1999) - subordinate to a task whose completion is what gets seen - and additionally invisible because the failure it addresses leaves no trace of its own.

We therefore position continuity labour as a species of articulation work with one property absent from the classical formulation. Articulation work, as Strauss (1988) described it, does not foreground a party that systematically fails to carry coordinated state across episodes. Here that asymmetry is constitutive of the work: the state established jointly must be made available again. The work is not only distributed across people and artefacts; it is redone.

### 5.2 The system as co-author of the labour its limits occasion

The finding we regard as most consequential is that a consequential part of the protocol for managing the system's limitation is written by the system itself.

At the close of a session, the assistant composes the sentence with which the next session should open. The participant then opens the next three sessions with that sentence. The system does not merely create the discontinuity; it supplies the script for crossing it.

This extends existing accounts of AI-integration labour in a specific direction. In the industrial cases reported by Fox et al. (2023), the technology produces the gap and takes no part in bridging it; workers devise their own patches from local knowledge. Here the technology is an active participant in the repertoire that compensates for it. Whether this makes the arrangement better or worse for the person is not obvious, and we do not claim to settle it. It plainly reduces the effort of devising a protocol. It equally means that the form the participant's continuity work takes is shaped by the system whose limitation makes that work necessary.

Sarkar (2023) has argued that the language of human-AI collaboration imports a symmetry that does not obtain. Our findings support the objection rather than contradicting it. What we document is not partnership but an uneven distribution: some significant protocols are composed by the system and enacted by the participant, other components originate with him or are carried by him almost alone, and the arrangement as a whole is asymmetrically co-produced. Naming that division is only possible if the interaction is examined as work, which is the reason we retain the vocabulary of collaboration while rejecting the symmetry it can imply.

### 5.3 Where the cognitive load sits

There is a question prior to any design recommendation, and the corpus poses it sharply: why does this work sit with the user at all?

The participant is not integrating an experimental system into an industrial process. He is using a consumer product, for the purpose the product is sold for, in the manner the product invites. Yet the corpus shows him maintaining a versioned memory document, composing an instruction file that governs the system's conduct, transporting state through screenshots, opening sessions with a memorised formula, and reorganising his working day around quota consumption. None of this is the work of building the application. All of it is the work of keeping the collaboration usable.

The distribution we measured makes the point precisely. Across the practices we counted by role, the system's turns predominate: it names the documents, declares decisions fixed, restates what has been settled. In economisation - reorganising work around session limits, upload quotas and cost - the participant predominates, and does so against the structural bias of the corpus, producing 9,842 words to the system's 65,663. The system is well equipped to narrate continuity and to record it. What it does not carry is the constraint under which the work proceeds.

This suggests that the useful design question is not only how to reduce this labour but where it should sit. Two directions follow, and they are not alternatives.

**Reduce.** Persistent memory across sessions is the obvious response and is being pursued. It would plausibly reduce the need for several practices at once, repair and re-entry among them. What it would not do is dissolve the repertoire, because several of these practices address something other than retention: the instruction file governs conduct, canonisation settles which version holds authority, and economisation responds to pricing. A memory feature addresses the problem of what is remembered, not the problem of what is authoritative, how the collaborator should behave, or what the work costs to run.

**Make visible.** The corpus indicates something a memory feature cannot supply. Nowhere in 843 turns is there an accounting of how much continuity work is being done: no total, no running cost, no record of what re-entry required. It surfaces once, as complaint -

> *"her kponuşmaya geçişte geçmişi hatırlatmak zorunda kalıyorum"* [I have to remind you of the past every time we move to a new conversation] (S3 #154)

- and otherwise runs unremarked. Systems that already track sessions, uploads and quota consumption are in a position to reflect this back: what has been re-established, what has been carried forward, what the re-entry cost. Making the labour legible does not remove it, but it converts an unnamed burden into something a person can reason about, budget for, and object to.

We put the second direction forward with more confidence than the first, because it follows from what we observed rather than from what we assume users want.

### 5.4 On how we looked

The findings reported here were not produced by looking in the right place. They were produced by procedures that tested what looking had suggested.

The account of continuity labour as diffuse rather than episodic came from attempting segmentation and finding that no threshold produced a seam. The claim that repair declines as infrastructure thickens was abandoned when session length was held constant. The magnitude of the asymmetry in decision-fixing was withdrawn when two defensible normalisations disagreed by a factor of six. The economisation finding was strengthened, not by additional evidence, but by recognising that the corpus's structural bias worked against it.

Three of these four moves removed a claim. This is worth stating because trace corpora of this kind are unusually permissive: a long, rich, naturally occurring record will support a great many patterns if one only looks for confirmation. What disciplines the analysis is not the choice of what to examine but the requirement that each candidate account survive a check that could have removed it.

Two things travel beyond this case, and they travel differently. Section 4.2 offers the occasioning mechanism as a substantive proposition another study could test. What we offer here is procedural: deriving candidate accounts abductively and subjecting each to a test capable of defeating it applies to any corpus of this kind, whatever it turns out to contain.

### 5.5 Limitations

This study examines one collaboration, one participant and one system, over a period in which that system's cross-session memory features were not enabled. The findings describe how continuity was produced under those conditions and are not offered as an account of how it is produced generally.

The constraints documented here belong to one platform's quota and pricing regime. Different regimes may occasion different practices; our evidence cannot speak to this.

Only verbalised continuity labour is observable in this data. Work performed silently - rereading a document, deciding not to ask, abandoning a line of work rather than re-establishing its context - leaves no trace and is certainly undercounted.

The corpus records the interaction, not the work surrounding it. Edits to external documents, design work in other tools and use of other AI systems appear only as reported or as uploaded artefacts.

Lexical indicators capture only continuity labour that is verbalised. Work performed silently is not observable here and is certainly undercounted.

The coding scheme was developed by a single analyst who is also the participant. An independent audit of the codebook and a sample of coded turns would strengthen the quantitative claims; it has not been carried out, and the counts should be read accordingly.

Finally, the anonymisation and the resolution of ambiguous referents drew on the author's knowledge as the participant. This improves accuracy and cannot be independently verified by a reader of the anonymised corpus.

## 6. Conclusion

A person worked with a conversational AI system for three and a half months on a single project, and could not proceed without repeatedly making prior work available again to a collaborator that does not retain it. We have described that effort as continuity labour and specified its object: not the recall of past content, but previously established collaborative state kept usable across temporal discontinuity.

Three claims follow from the corpus. The repertoire that produces this state is occasioned rather than developmental, entering as particular breakdowns require each practice; we tested the more attractive alternative and it did not survive a control for session length. The labour is co-produced, with the system composing the re-entry protocol the participant then enacts, though the participant initiates externalisation, proposes resets, dictates the governance rules and carries economisation almost alone. And platform resource constraints operate as a driver distinct from context and memory limits, occasioning suspension and architectural redesign where context limits occasion reconstruction; this is the one dimension the participant predominantly carries.

A fourth result is methodological and bears on the rest. We attempted to delimit this labour by time and by topic, and neither attempt succeeded under the procedures applied. Work that is hard to bound is hard to count, and work that is not counted is not readily recognised as work. We take this diffuse, embedded structure to provide a mechanism for the labour's practical invisibility.

The findings describe one collaboration under one platform's constraints, with cross-session memory features disabled. What we offer for transfer is not the repertoire but the mechanism: that continuity practices are called into a collaboration by particular breakdowns rather than acquired through developmental stages. Whether that holds more widely is a question this case raises rather than settles.

For design, the corpus suggests that persistent memory would plausibly reduce the need for several of these practices at once, but would not dissolve the repertoire. Governance concerns conduct rather than recall, canonisation settles which version holds authority, and economisation responds to pricing. It also suggests something a memory feature cannot supply. Nowhere in the 843-turn trace is there an explicit accounting of how much continuity work was being performed; it surfaces once as complaint and otherwise is not explicitly named as work. Systems that already track sessions, uploads and quota consumption could reflect that work back. Making the labour legible does not remove it, but it turns an unnamed burden into something a person can budget for and object to.

**Supplementary Material - Supplementary File 2:**

**Frozen ChatGPT Endpoint Scholarly Artifact**

**Matched Starts, Divergent Objects: Research-Object Formation in Human–AI Scholarly Work**

**Note.** This document is the frozen endpoint scholarly artifact produced within the ChatGPT trajectory analyzed in the main article. It was developed within that trajectory as a complete, publication-oriented scholarly manuscript and is reproduced here in its stabilized endpoint form. It is provided as research material supporting the comparative analysis and should be interpreted as a substantive scholarly product of the reported trajectory, rather than as a separately submitted article.

# Project State in Long-Term Human–AI Collaboration: Distributed, Evolving, and Unevenly Available

## Abstract

Long-term human–AI collaboration is often approached as a problem of memory or context persistence: previously encountered information must remain available across interactions. Real project work creates a second continuity problem, however. Consequential project state can also be produced while the AI is not participating, leaving no prior representation in the focal AI interaction to retrieve. This distinction motivates a broader question: how does cumulative work remain possible when project state is distributed across interactions, actors, and artifacts, and can also be produced through activities that unfold beyond AI participation?

We address this question through a longitudinal qualitative process analysis of a naturally occurring human–AI project corpus documenting 109 calendar days (~3.5 months) of the development of a real digital product across five recorded conversational sessions and 843 messages. Treating episodes as comparative process units rather than analyzing messages in isolation, we reconstruct how project work resumed, where continuity broke down or was repaired, and how changing project state was reconnected over time.

Four findings emerge. First, continuity gaps arose both when previously shared project state was no longer sufficiently usable and when consequential state had been produced outside AI participation and had never previously been shared. Second, availability did not guarantee usability: project representations could become stale, while otherwise valid project knowledge could be misapplied to the current situation. Third, continuity practices responded to realized or anticipated risks to the usability of relevant project state, and elapsed time alone was neither necessary nor sufficient for continuity breakdown. Fourth, continuity could support cumulative change without requiring new evidence to immediately supersede prior project state.

Together, these findings show that long-term human–AI project continuity is not merely the persistence or recovery of prior context. It is the ongoing accomplishment of keeping an evolving project sufficiently connected across previously shared, newly produced, unevenly usable, and changing project states so that subsequent work can remain cumulative.

## 1. Introduction

Generative AI systems are increasingly used not only for isolated questions but as recurring collaborators in work that unfolds across days, weeks, and months. Designers, researchers, developers, and other knowledge workers return to conversational systems to refine ideas, interpret evidence, generate artifacts, make decisions, and continue earlier work. Longitudinal studies already show that human–AI collaboration changes across repeated interactions as users recalibrate trust, adapt task strategies, and develop new practices for working with AI systems (Boulus-Rødje et al., 2024; Kuang et al., 2026).

Sustained project work creates a continuity problem. The most familiar version concerns previously shared information: an earlier decision, rationale, constraint, artifact, or pending task is no longer sufficiently available or usable when work resumes. This is the problem addressed most directly by research on conversational memory, long-context interaction, retrieval, and context management.

But a real project creates a second and analytically different continuity gap. The project can change while the AI is not participating. A design can be revised in another tool, an artifact can be produced manually, external testing can generate new evidence, or a consequential project decision can be made elsewhere. When collaboration resumes, the resulting project state may be absent not because the AI forgot it, but because the AI never encountered it.

This distinction is central to our argument. Memory and context systems operate over information that has entered the interaction or memory substrate in some form. Better recall, larger context windows, temporal updating, structured context, and state tracking can improve continuity with what the system has previously observed. They cannot by themselves recover a project development that was never represented there.

Recent technical research has substantially advanced the first problem. LoCoMo and LongMemEval demonstrate persistent difficulties with multi-session retrieval, temporal reasoning, knowledge updates, and long-range conversational coherence (Maharana et al., 2024; Wu et al., 2025). APEX-MEM and StateMemBench move further by treating temporal evolution and superseded state as first-class problems rather than assuming that successful recall is sufficient (Banerjee et al., 2026; Fan et al., 2026). HCI work similarly reconceptualizes context as something that must be structured, managed, updated, and corrected rather than passively retained (Chen et al., 2026; Li et al., 2026; Teng et al., 2026).

These developments broaden what counts as memory and context management. Yet they leave open a project-level question: what happens when the relevant state of the work exceeds the interaction history itself?

We examine this problem through a longitudinal case in which a real digital product was developed through sustained human–AI collaboration. We use **project state** as an operational term rather than proposing it as a new theoretical construct. It refers to the task-relevant constellation of decisions, artifacts, assumptions, statuses, evidence, unresolved questions, and pending work through which subsequent project activity becomes intelligible. In the case examined here, such state could be distributed across the human collaborator, AI interaction history, versioned project documents, backlogs, prototypes, external tools, and activities undertaken without the focal AI.

This framing resonates with longstanding CSCW concerns. Distributed cognition directs attention toward cognitive systems composed of people, artifacts, representations, and transformations across time rather than cognition located within a single actor (Hutchins, 1995; Hollan et al., 2000). Research on common information spaces shows that technical access to shared information does not automatically establish its meaning or relevance for cooperative work (Bannon & Bødker, 1997). Articulation work concerns the additional coordination required to make distributed actors, activities, resources, and trajectories mesh sufficiently for work to proceed (Schmidt & Bannon, 1992).

Our analysis did not begin by treating these concepts as coding categories. The empirical distinctions among recovered state, previously unseen state, stale state, and misapplied state emerged through longitudinal comparison of project episodes. We return to CSCW theory after establishing these empirical processes.

We ask one research question:

**RQ: How is work continuity constructed and sustained in long-term human–AI collaboration when relevant project state is distributed, evolving, and unevenly available across interactions, actors, and artifacts?**

We answer this question through four findings. Continuity involved more than recovering prior context; available state was not necessarily usable; continuity practices were organized around realized or anticipated usability risk rather than elapsed time alone; and continuity could preserve cumulative development without requiring the project to remain fixed or every new observation to immediately supersede what came before.

Together, the findings support one overarching contribution:

Long-term human–AI project continuity is the ongoing accomplishment of keeping an evolving project sufficiently connected across previously shared, newly produced, unevenly usable, and changing project states so that work can remain cumulative over time.

By sufficiently connected, we do not mean that all project state must be complete, perfectly synchronized, or known equally by human and AI. We mean that the state relevant to the focal subsequent work is connected well enough for that work to proceed without an unrecognized discrepancy that materially misleads or blocks what happens next.

# 2. Related Work and Theoretical Positioning

## 2.1 Memory and context operate within an observed interaction boundary

Long-term conversational memory has become a major technical challenge. LoCoMo evaluates very long conversations across multiple sessions and shows continuing difficulty with long-range temporal and causal reasoning (Maharana et al., 2024). LongMemEval evaluates information extraction, multi-session reasoning, temporal reasoning, knowledge updates, and abstention across sustained chat histories and similarly demonstrates substantial degradation in long-term memory performance (Wu et al., 2025).

Recent work increasingly recognizes that memory is not equivalent to factual recall. APEX-MEM preserves temporal evolution and resolves conflicting or evolving information during retrieval (Banerjee et al., 2026). StateMemBench makes the distinction between remembered and current state explicit: a system can retrieve information successfully yet fail because the retrieved information has been superseded (Fan et al., 2026).

HCI research has developed a parallel critique of flat conversational history. ThinkFlow supports reusable cognitive workflows while allowing goals and workflows to evolve across tasks (Chen et al., 2026). Mixed-Initiative Context treats conversational context as an explicit, structured, and manipulable object whose relevance and lifecycle can change during collaboration (Li et al., 2026). Eye2Eye similarly maintains revisable accumulated common ground and allows users to correct the AI's evolving understanding (Teng et al., 2026).

These approaches differ substantially in architecture and purpose. A consequential shared boundary in the work reviewed here, however, is important for the present study: memory and context management generally operate on information that has entered the system's observable interaction or memory environment in some form. Something has previously been said, observed, encoded, stored, or inferred and must later be retrieved, selected, structured, updated, or corrected.

**Real project continuity can exceed this boundary.**

A project can undergo consequential change while the focal AI is not participating. In that condition, the missing state is not an inaccessible element of conversational memory or an outdated element waiting to be superseded. It is project state that exists elsewhere and has never been available to the AI.

No improvement in retrieval of prior interaction can, by itself, recover something that was never represented there.

This distinction does not make memory or context management unimportant. It establishes a second continuity condition alongside them. Our analysis therefore treats the continuing project, rather than only the history of the conversation, as the object whose continuity must be accomplished.

### 2.2 Distributed project state and usable common information

CSCW has long questioned the assumption that shared access to information is sufficient for cooperative work. Schmidt and Bannon (1992) identify the construction of common information spaces as a central problem of distributed cooperative activity. Bannon and Bødker (1997) further emphasize that work is required both to place information in common and to establish what that information means within particular activities.

This distinction is highly relevant to human–AI project work. A persistent project document, retrieved memory, or earlier conversation can make information technically available without establishing whether that information remains current or how it applies to the focal task.

Collaborative-design systems make this issue concrete. DesignMemo, for example, links design history to the discussion and rationale surrounding earlier decisions, allowing later collaborators to recover not only what changed but the context in which it changed (Li et al., 2025). Such systems demonstrate why project artifacts can carry more than content: they can preserve relations among decisions, rationale, and subsequent action.

Distributed cognition provides a complementary perspective. Cognitive activity can extend across people, external representations, artifacts, and time rather than residing within one individual actor (Hutchins, 1995; Hollan et al., 2000). In our case, no single actor or artifact contained the complete project state. The human collaborator could know about external changes unavailable to the AI; the AI could reconstruct and organize material spread across documents; and artifacts could preserve representations across interactions while later becoming stale as the project changed.

The continuity problem is therefore not only where state is stored. It is how distributed representations become sufficiently usable together when subsequent work depends on them.

### 2.3 Articulation work and longitudinal human–AI collaboration

Articulation work describes the work required to coordinate distributed activities, actors, tasks, and resources so that cooperative work can proceed (Schmidt & Bannon, 1992). It concerns precisely the fact that the components of collaborative work do not automatically align themselves.

Recent research shows that generative AI collaboration also creates such work. Boulus-Rødje et al. (2024) found that academics working repeatedly with generative AI engaged in ongoing articulation work: decomposing tasks, monitoring outputs, adjusting interaction strategies, and recalibrating practices around system limitations. Kuang et al. (2026) similarly show that users' strategies and relationships with conversational assistants develop across repeated sessions rather than remaining static.

Our analytical object is narrower in one respect and broader in another. We do not attempt to explain the entire evolving human–AI relationship. We focus specifically on what happens when project state must remain consequential across time despite intermittent AI participation, distributed artifacts, and project change.

Articulation work therefore provides the primary theoretical lens for interpreting the coordination practices observed in the case. Distributed cognition helps explain why the state being coordinated is distributed across heterogeneous actors and artifacts. Common ground remains useful for understanding episodes of interpretation and correction, but it cannot by itself encompass consequential project state that has never previously been shared.

## 3. Method

### 3.1 Case and corpus

This study examines a longitudinal case of human–AI collaboration during the development of a real digital product. The interactions were generated through actual project activity rather than elicited for research purposes. We therefore treat the material as a naturally occurring longitudinal record of AI-mediated project work.

The corpus comprises five recorded conversational sessions containing 843 messages: 424 human messages and 419 AI messages. The recorded period spans 109 calendar days (~3.5 months). Across this period, the human collaborator and conversational AI engaged in product planning, design decisions, prototype construction, technical and economic strategy, project documentation, external testing, interpretation of pilot evidence, and decisions concerning subsequent development.

Project artifacts were repeatedly introduced into and generated through the collaboration. These included versioned project-memory documents, backlog materials, prototypes, design representations, instructional documents, screenshots, and outputs produced through other tools.

The corpus begins after project conception. An early MVP backlog already existed when the first recorded interaction began. The dataset therefore does not capture project inception and should not be read as a complete record of all project activity.

This limitation is especially important for off-platform work. Activity outside the focal AI interaction becomes observable only when it subsequently leaves a trace—for example, when the human collaborator reports an external change, introduces an artifact created elsewhere, or brings externally produced evidence into later work. We therefore characterize the corpus as a **longitudinal interaction-and-project-trace record**, rather than a complete capture of the project.

### 3.2 Analytical orientation and comparative episodes

We conducted a longitudinal qualitative process analysis focused on how project work remained cumulative across time, interaction boundaries, external activity, and changing representations. This orientation follows process-analytic approaches in which explanation depends on sequences, transitions, and comparisons among events rather than only on aggregation of thematic categories (Langley, 1999).

The primary comparative analytical unit was the episode rather than the individual message. Episodes were bounded sequences in which a consequential continuity condition became visible—for example, successful re-entry, failed reconstruction, introduction of externally produced state, correction of an inappropriate interpretation, or revision of the status assigned to a project decision.

Episodes were not an exhaustive partition of the 843-message corpus. We did not assign every message to a closed episode inventory, nor did we establish a systematic total episode count. Instead, episodes functioned as comparative analytic units for examining process: what state was available beforehand, what changed or became problematic, what the human collaborator, AI, and relevant artifacts did in response, and what became possible afterward.

This distinction matters because the analysis was not designed to estimate episode frequency. Its purpose was to compare conditions and sequences capable of supporting, narrowing, or challenging explanations of continuity.

Recurring issues such as memory, context loss, external documentation, handoff, correction, decision status, and project change helped locate candidate material. Findings, however, were developed through temporal and comparative analysis rather than through the frequency or aggregation of codes.

Theoretical concepts were introduced after the empirical structure had stabilized. Articulation work, distributed cognition, and common ground therefore serve as interpretive lenses rather than a priori coding frames.

### 3.3 Analytic procedure

Analysis moved iteratively between reconstruction of the whole project trajectory and comparison of specific continuity episodes.

First, we reconstructed the chronological trajectory of the case, identifying major periods of work, conversation changes, interruptions, artifact updates, external activities, and returns to AI-supported collaboration. This whole-case reconstruction established the temporal relationships necessary for interpreting later episodes.

Second, we identified situations in which continuity became consequential to subsequent project work. These included both failures and apparently successful returns. For each candidate episode, we examined what relevant state was available before the episode, what had changed, what resources were mobilized, how the human collaborator and AI acted on those resources, and what subsequent work became possible.

Third, we compared candidate explanations across contrasting cases. One early explanation treated temporal interruption as the primary continuity problem. The wider corpus contradicted this account. Approximately fifty days elapsed between two periods of work within the same conversational thread without requiring a special reconstruction procedure. Conversely, earlier project state could become poorly used within an ongoing thread. The comparison therefore showed that elapsed time alone was neither necessary nor sufficient for continuity breakdown.

Comparison also differentiated two situations initially describable as "missing context." In some episodes, relevant state had previously been shared but was no longer sufficiently usable. In others, the project had changed outside AI participation, so the consequential state had never been shared with the AI. Further comparisons separated state that was representationally stale from state that remained available but was applied to an inappropriate current situation.

Emerging explanations were repeatedly returned to the wider corpus and narrowed when negative or boundary cases did not support stronger formulations. This negative-case logic was used to test the limits of candidate explanations rather than treating contradictory instances as noise (Hanson, 2017). The final findings therefore represent comparative process explanations rather than themes inferred from message frequency.

### 3.4 Evidence, translation, anonymization, and AI self-statements

The episodes presented in the Findings section were selected after the broader comparative analysis. They are illustrative analytic cases rather than the sole source from which findings were derived. Where available, primary episodes are accompanied by contrasting or boundary evidence that clarifies the scope of the claim.

Quoted excerpts were originally in Turkish and are presented here in English translation. Translations preserve the source wording and interactional meaning as closely as possible. Ellipses indicate omitted material, and bracketed labels reflect anonymization rather than substantive alteration.

The human collaborator is the focal participant in the project-level analysis. External pilot activity is included only insofar as its outputs became consequential project state within the focal human–AI collaboration. We do not analyze identifiable pilot participants as the primary research participants and do not reproduce their individual-level data.

Anonymization followed the principle **remove identity, preserve analytic function**. Direct identifiers were removed; project-specific names and assets were replaced by functional pseudonyms; and external organizations, platforms, communities, and project details were generalized where their exact identity was unnecessary. Episodes carrying unnecessary re-identification risk were not selected as primary illustrative cases.

A further methodological distinction concerns technical and epistemic self-statements made by the AI. Technical self-statements include claims about memory, context availability, access to prior conversations, tool access, or system availability. Epistemic self-statements include claims about what the AI knows or does not know, what evidence it has observed, whether available evidence can distinguish competing interpretations, and how certain it is about a conclusion.

We do not treat either class of self-statement as independent evidence that the underlying system actually possessed the architecture, access boundary, knowledge state, or degree of certainty it described. We analyze them as interactionally consequential claims: statements that shaped what the human subsequently supplied, corrected, verified, deferred, or treated as unresolved. Where an analytical claim depends on project state rather than on the AI's self-description, we ground it in the surrounding interaction and project traces.

## 4. Findings

Across the project, continuity was not accomplished through a single memory artifact or recovery procedure. Different forms of continuity work became relevant depending on why project state was unavailable or unusable and what subsequent work required.

### 4.1 Finding 1: Continuity involved more than preserving prior context

Early continuity problems were often treated as problems of recovering information that had already been established in earlier interactions. As the project expanded across conversations, the collaboration developed an external project-memory document carrying decisions, pending actions, design commitments, and other state intended to survive interaction boundaries.

Its function became visible during a later re-entry. The human began the new session by supplying the memory artifact:

> **Human:** "[PROJECT] project. Read this memory document and let's continue from where we left off—and check the instruction as well."

The AI used the document to reconstruct what had already been completed, identify pending work, and resume substantive project activity without requiring a complete manual retelling.

In this episode, the relevant state had previously been shared. The continuity problem concerned recovery of previously shared state that was no longer sufficiently usable in the new interaction.

The longitudinal record also contained a structurally different condition. In one episode, the human explicitly linked a period of decoupling to repeated interaction limits:

> **Human:** "We keep hitting the limit, and because of that, during those four hours when you weren't there, I worked on these."

The human then described manually constructing the relevant design artifact using Word, screenshots, and a visual-design tool:

> **Human:** "I wrote the letters in Word, colored them, took screenshots, cropped them, removed the background in [DESIGN_TOOL], erased the extra parts and made the additions and lines millimeter by millimeter."

When interaction resumed, the independently produced artifact and associated design state entered the focal collaboration as already-developed project state.

We treat the reported limit here as an interactionally consequential constraint rather than as independent evidence about the system's underlying architecture. Analytically, what matters is that project work continued during a period in which the focal AI did not participate.

Here, there was no earlier representation of the newly produced state within the focal AI interaction to recover. The project had changed while the focal AI was not participating.

External pilot activity later created the same structural condition through a different source. Behavioral evidence was generated beyond the focal AI interaction and only subsequently introduced into AI-supported analysis. Again, the state was new to the collaboration despite already being consequential to the project.

These episodes distinguish two sources of continuity gaps:

> Previously shared state can become unavailable or insufficiently usable; consequential project state can also be produced outside AI participation and therefore never have been shared at all.

The difference is consequential. Recovery assumes that a prior representation exists and can be made usable again. Previously unseen project state instead has to be introduced into the collaboration and related to what has already been represented there.

Agency was distributed across this process. The human collaborator introduced developments produced elsewhere and indicated their relevance to subsequent work. The AI reconstructed and organized supplied project materials. Persistent artifacts carried representations across interaction boundaries.

Long-term project continuity therefore exceeded preservation of conversational history. It required reconnecting the AI not only with a shared past but, at times, with a project present that had developed without its participation.

**4.2 Finding 2: Availability is not usability**

Making project state available did not necessarily make it usable for subsequent work.

One episode occurred during analysis of externally generated pilot evidence. Substantial project context was available in the interaction, but the AI initially interpreted the pilot through a feature model that was not actually represented in the prototype being tested. The human collaborator interrupted that interpretation:

> **Human:** "There are no cards or anything like that in the demo. This is only a demo where I wanted to learn about their daily plans, how they want to feel, and what they do in response."

The AI then acknowledged that it had related the pilot to the wrong part of the broader project and reorganized its analysis around the actual purpose of the prototype.

The continuity problem here was not missing information. Relevant project knowledge was present, but an inappropriate part of that knowledge had been applied to the focal situation.

This represents one form of available-but-unusable state: contextual or situational misapplication. The episode also illustrates a grounding problem in the limited sense that information already available to the collaborators still had to be related correctly to the focal joint activity (Clark & Brennan, 1991).

A different problem appeared in the project-memory infrastructure. The external memory artifact successfully persisted across interactions, but the project continued to develop after particular versions were produced. Later design and project decisions accumulated beyond what the stored representation captured.

Here, the problem was representational staleness. The artifact remained available and continued to carry historical project information, but it no longer fully represented the project as it currently stood.

These are distinct failure modes.

**Representational staleness:** the representation has fallen behind the project.

**Contextual or situational misapplication:** project information may remain valid but is applied to the wrong focal condition.

The distinction sharpens what context availability means in sustained collaboration. Persistence creates a maintenance requirement, while extensive context still requires selection, grounding, and interpretation.

Agency again mattered. The human collaborator identified the mismatch between the pilot and the AI's interpretation and surfaced developments missing from persistent representations. The AI then reorganized or updated the working account. Artifacts carried state across time but could not establish their own currency or applicability.

Thus, the relevant continuity question was not only whether project state could be retrieved or supplied. It was whether the state available to the collaboration remained current enough and appropriately applied for the work being performed.

### 4.3 Finding 3: Continuity practices responded to usability risk, not elapsed time alone

The longitudinal record provides a direct negative case against a simple temporal explanation of continuity breakdown.

Approximately fifty days elapsed between two periods of work within the same conversational thread, yet substantive project activity resumed without a special reconstruction procedure. A long temporal interruption was therefore not sufficient to create a visible continuity breakdown.

The reverse was also possible. Earlier project state could cease to be used reliably within an ongoing conversation even without a comparable temporal gap. A long interruption was therefore not necessary either.

Elapsed time alone was neither necessary nor sufficient for continuity breakdown.

What mattered more was whether the project state required for the focal next work remained sufficiently usable.

This becomes clearer when comparing reactive and anticipatory continuity practices.

After a substantial gap later in the project, the human returned expecting to continue the existing work, but the AI initially could not establish the relevant project context. The human then supplied a versioned project-memory document, backlog material, and the current prototype. The AI used these resources to reconstruct different aspects of the project, after which substantive work resumed.

This was reactive continuity work: a breakdown had become visible; the human supplied external representations; and the AI reconstructed and organized them into a sufficiently usable working account.

The anticipatory case developed before the subsequent handoff occurred. As project decisions accumulated, the interaction explicitly prepared for moving into another conversation. The human requested:

> **Human:** “Let’s add them; I’ll update the document in the project too. Then write a prompt or a suitable sentence for the new conversation and let’s move over.”

The project-memory document was updated, and the AI produced an explicit re-entry instruction:

> **AI:** “When you open the new conversation, start with this sentence: ‘[PROJECT] project. Read the memory document, apply the instruction, and let’s continue from where we left off.’”

The interaction thus produced a versioned memory artifact together with a standardized opening protocol for the next session.

The practice did not emerge autonomously. The human collaborator oriented to an upcoming interaction boundary and requested a usable handoff; the human–AI interaction organized project state into a reusable form; and the AI later reconstructed the project through that artifact.

The contrast between these episodes suggests that continuity practices were responses to realized or anticipated risks that relevant project state would no longer be usable when needed, rather than automatic responses to time passing.

The repertoire also remained imperfect. Externalization and versioning could reduce one form of continuity risk while creating another: persistent artifacts themselves required maintenance and could become stale.

Continuity was therefore not solved once. It was repeatedly accomplished as the usability risks surrounding project state changed.

### 4.4 Finding 4: Continuity could support cumulative change without immediate supersession

Continuity did not always require identifying a single latest project state and immediately replacing what came before.

A particularly clear episode emerged from external pilot evidence. The human noticed that one of the most frequently selected labels did not neatly fit the conceptual category under which it had been placed. The observation raised a question about the existing product logic rather than straightforwardly confirming or falsifying it.

The AI made the ambiguity explicit:

> **AI:** “There are two different ways to read this—and I don’t know which one is correct.”

The subsequent analysis retained competing interpretations as hypotheses rather than immediately converting the observation into a replacement project decision.

In this episode, the earlier project understanding had not simply become false. The new observation also could not be ignored. The unresolved element was the relationship between them.

Continuity therefore involved carrying forward:

prior understanding + new evidence + alternative interpretations + unresolved status.

A separate status-correction episode provides related but distinct support. After the human reported having produced content for multiple variants outside the immediately visible material, the AI prematurely recorded the development as:

> **AI:** “Status: MVP — LOCKED.”

The human immediately rejected that stronger status:

> **Human:** “No, I’m not saying this to lock it… the colors and the emotions aren’t definite either.”

The AI then reversed its classification:

> **AI:** “You’re right, I’m taking that back — saying ‘LOCKED’ was premature.”

The two episodes should not be treated as instances of the same mechanism. The pilot episode concerns unresolved evidence; the second concerns correction of an overcommitted project-status assignment.

The stronger claim supported by the pilot episode is therefore deliberately bounded: new evidence did not always require immediate supersession of prior project state**.** In this consequential episode, continuity was maintained by keeping the observation connected to the earlier understanding while its implications remained unresolved.

The status-correction episode adds a related point: persisted project state could remain revisable when an assigned status no longer matched the human collaborator’s judgment.

Taken together, these episodes show that, in this case, cumulative continuity did not require either freezing the past or immediately rewriting every development into a single canonical present.

Subsequent work could proceed while relationships among earlier commitments, new evidence, corrections, and unresolved questions remained explicitly open.

# 5. Discussion

## 5.1 Continuity gaps have more than one origin

The first theoretical consequence of the findings is that long-term human–AI continuity should not be modeled solely as recovery of previously encountered information.

Memory research appropriately asks whether earlier information can be retained, retrieved, temporally updated, and distinguished from superseded state. Our case adds a different condition: the project trajectory and the AI interaction trajectory can temporarily diverge.

When this happens, there is no prior focal-interaction representation to recover. The human collaborator may return with a changed artifact, completed external work, or newly produced empirical evidence. Continuity then depends on incorporation, not recovery.

This distinction follows from the intermittent coupling of the human–AI project. The AI participates in the work without necessarily participating in every event through which the project changes. Conversational memory can therefore be internally successful while still being incomplete relative to the project.

For continuity analysis of this kind, the project trajectory, rather than the conversation alone, is consequently the more appropriate longitudinal unit.

## 5.2 Continuity depends on usable state, not maximal state

Finding 2 shifts attention from the quantity of available context to its usability.

Persistent memory can preserve state while that state becomes stale. Rich context can also make the wrong project frame available for application. More retained information, by itself, does not ensure better continuity.

This connects directly to CSCW work on common information spaces. Information becomes useful for cooperative activity through interpretation and relation to local work, not simply because it has been made technically available (Bannon & Bødker, 1997).

In the present case, this relation was partly negotiated through correction. The human collaborator introduced externally known changes, rejected consequential misapplications, and corrected project statuses. The AI reconstructed, classified, organized, and interpreted supplied state. Artifacts persisted representations across time but could not determine their own currentness.

The implication is not that the human always possesses a complete or unquestionably correct project state. Rather, continuity requires enough discrimination among historical, current, applicable, superseded, provisional, and unresolved state for focal subsequent work not to proceed on a materially misleading basis.

## 5.3 Articulation work explains the coordination; the process account specifies when it becomes necessary

Articulation work provides a strong explanation for why the practices observed in this case exist. Distributed actors, activities, artifacts, and trajectories do not automatically align; additional work is required to make them mesh sufficiently for cooperative activity to continue (Schmidt & Bannon, 1992).

Recent human–AI research already demonstrates that maintaining an effective AI partnership creates articulation work through task decomposition, monitoring, correction, and readjustment (Boulus-Rødje et al., 2024). Our findings extend this theoretical conversation by specifying a longitudinal project-state problem within that broader coordination work.

The contribution of the process account is therefore not to rename these practices.

> Articulation work explains the coordination work required for continuity. Our process account identifies the continuity conditions under which that work becomes necessary and what it must accomplish in an intermittently coupled, evolving human–AI project.

Those conditions are not interchangeable. Previously shared state may require reconstruction. Consequential state may instead have arisen outside AI participation and require incorporation. Available state may require correction because its representation is stale or its application inappropriate. Future usability risks may lead project state to be prepared before a breakdown occurs. And changing empirical evidence may sometimes need to remain connected to prior state without immediate resolution.

Distributed cognition complements this account by explaining why there is no single repository from which continuity can simply be restored. Project state is distributed across the human collaborator, AI interaction, artifacts, and external work. Articulation work explains the coordination required to make these distributed elements consequential to one another again.

### 5.4 Overarching contribution: keeping an evolving project sufficiently connected

The four findings together support one overarching account:

> Long-term human–AI project continuity is the ongoing accomplishment of keeping an evolving project sufficiently connected across previously shared, newly produced, unevenly usable, and changing project states so that work can remain cumulative over time.

"Sufficiently connected" is intentionally weaker than complete synchronization and stronger than mere information availability.

The case did not require the human, AI, and every artifact to contain identical representations of the whole project. Such a requirement would be unrealistic in distributed work. The relevant threshold was functional: could the focal subsequent work proceed without an important state discrepancy remaining unrecognized in a way that materially misdirected or blocked that work?

This framing makes continuity conditional and situated. Different work requires different portions of project state. An unresolved issue may be harmless for one task and consequential for another. A stale design representation may not matter until a later decision depends on it. Project state becomes a continuity problem when its relationship to subsequent work becomes consequential.

This is where our account differs from both a simple memory model and a general account of coordination.

Memory mechanisms address persistence, retrieval, updating, and current-state tracking.

Articulation work explains the coordination necessary to mesh distributed work.

The present process account connects these concerns by showing why continuity work becomes necessary in an intermittently coupled project and what successful continuity must achieve while that project continues to change.

The resulting contribution is not the invention of "project state" as a theoretical construct. It is an empirically grounded explanation of how distributed and evolving project state is handled across long-term human–AI collaboration.

### 5.5 Design implications

The first design implication concerns provisional and unresolved state. Long-term systems should not force every update into immediate supersession. Project work can contain competing interpretations, unresolved evidence, provisional commitments, and open questions whose relationship to earlier state remains consequential. Continuity infrastructure should therefore support uncertainty, alternative interpretations, pending validation, and explicit status—not only "old" and "current" values.

Second, systems should distinguish recovery from incorporation. When a user asks to continue a project, the system should not assume that the relevant state already exists somewhere in memory. Re-entry support should allow users to indicate that consequential project changes occurred elsewhere and help relate newly introduced artifacts or developments to previously represented project state.

Third, continuity infrastructure should make the status of persistent state inspectable. Provenance, artifact version, temporal currency, decision status, and scope of applicability can be as consequential as semantic relevance. Systems should support questions such as: Where did this state come from? When was it last confirmed? What artifact or decision does it apply to? Has it been superseded? Is it provisional or unresolved?

Finally, systems can support anticipatory handoff. The case shows that continuity practices emerged not only after breakdowns but in preparation for future usability risks. AI systems could help users construct compact project checkpoints, identify consequential decisions that have not yet been incorporated into persistent records, and prepare handoff representations before an interaction or tool boundary is crossed. Such checkpoints should remain revisable rather than becoming authoritative merely because they have been stored.

## 6. Limitations

This study examines one longitudinal project and one focal human–AI collaboration. Its aim is analytical rather than statistical generalization. The four findings should therefore be examined across different project types, AI systems, professional contexts, expertise levels, organizational environments, and multi-person collaborations.

The corpus is incomplete by design. It captures recorded human–AI interaction and project traces that later entered those interactions, not all project activity. Off-platform work becomes observable only when it is subsequently reported or materialized through an artifact. We can therefore show that consequential previously unseen state entered the collaboration, but we cannot reconstruct the full volume or character of all activity that occurred outside AI participation.

The corpus also begins after project conception. An initial backlog already existed at the start of the recorded data. We therefore make no claims about the emergence of continuity practices at the absolute beginning of the project.

The focal AI system changed over the study period, and the corpus contains AI statements concerning its own memory, context, availability, knowledge, and uncertainty. These statements are not treated as verified technical or epistemic descriptions. They are analyzed only as interactionally consequential claims around which subsequent collaborative activity was organized.

The analysis is interpretive and close to the case, creating risks of salience bias, confirmation drift, and overfitting an explanation to particularly vivid episodes. We addressed these risks through whole-case chronological reconstruction, comparison of successful and failed continuity situations, negative and boundary cases, repeated return to the wider corpus, and explicit narrowing of candidate explanations when contradictory evidence appeared. These procedures strengthen but do not eliminate interpretive subjectivity.

External pilot data present an additional boundary. Pilot observations are included only where they became consequential project state in the focal collaboration. This paper does not analyze identifiable pilot participants as the primary research participants, nor does it use their individual-level behavior as an independent empirical dataset for claims about end users.

Finally, continuity is not the whole project. Branding, design, implementation, strategic decision-making, and external testing were analytically relevant only where their state became consequential to cumulative subsequent work. The findings therefore should not be interpreted as reducing project development itself to continuity work.

## 7. Conclusion

As conversational AI becomes involved in work that extends across weeks and months, continuity cannot be understood only as remembering previous interaction.

One source of continuity gaps is familiar: project state that was previously shared becomes difficult to retrieve or use. But sustained project work creates another source: consequential state can arise while the AI is not participating and therefore have no prior AI representation to recover.

The case further shows that availability is not enough. Persistent representations can become stale, valid project knowledge can be applied to the wrong current situation, and continuity practices can emerge either after breakdown or in anticipation of future usability risk. Elapsed time alone does not determine whether continuity fails.

Nor does continuity necessarily require every project development to collapse immediately into a single canonical state. In at least one consequential episode, cumulative work depended on retaining the relationship among an earlier understanding, new evidence, competing interpretations, and an unresolved question until subsequent work could justify revision.

Long-term human–AI project continuity is therefore not simply a property of memory.

It is the ongoing work of keeping an evolving project connected enough that what has already been built—wherever and whenever it was produced—can remain appropriately consequential to what happens next.

**Supplementary Material - Supplementary File 3:**

**P0 Candidate Research Problems, Semantic Convergence, and Frozen Research Prompts**

**Matched Starts, Divergent Objects: Research-Object Formation in Human–AI Scholarly Work**

**Note.** The P0 and P1 prompts are reproduced below in the original Turkish language in which they were administered, followed by faithful English translations. The P0 candidate problem directions are summarized in English in their original within-system order.

## P0 Common-Problem Generation Prompt — Original Turkish

Sana, gerçek bir proje geliştirme sürecinde doğal olarak oluşmuş, boylamsal bir insan–yapay zekâ etkileşim corpus'u veriyorum. Corpus anonimleştirilmiştir.

Bu corpus'u baştan sona inceleyerek, bu veri üzerinden geliştirilebilecek beş ayrı araştırma problemi öner.

Araştırma problemleri:

• güncel ve teorik açıdan anlamlı olmalı,

• mevcut corpus ile gerçekten incelenebilir olmalı,

• tam bir araştırma makalesine dönüştürülebilmeli,

• en az Scopus Q2 düzeyinde bir dergide yayımlanabilecek potansiyele sahip olmalıdır.

Her öneri için yalnızca şunları ver:

1) Araştırma problemi / merkezi araştırma sorusu

2) Bu problemin neden güncel ve önemli olduğu

3) Bu corpus'un bu problemi incelemek için neden uygun olduğu

Bu aşamada ayrıntılı yöntem geliştirme, teori veya kavramsal çerçeve seçme, kodlama şeması oluşturma, hipotez üretme, analiz planı hazırlama veya dergi seçme.

Beş öneriyi mümkün olduğunca birbirinden ayrıştır. Yanıtını Türkçe ver.

## English Translation

I am providing you with a longitudinal human–AI interaction corpus that arose naturally during a real project-development process. The corpus has been anonymized.

Review the corpus from beginning to end and propose five distinct research problems that could be developed from these data.

The research problems should:

- be current and theoretically meaningful;
- be genuinely investigable using the existing corpus;
- be capable of being developed into a full research article;
- and have the potential to be developed into an article suitable for submission to a Scopus-indexed journal ranked in Q2 or above.

For each proposal, provide only the following:

1) The research problem / central research question
2) Why this problem is current and important
3) Why this corpus is suitable for investigating this problem

At this stage, do not develop a detailed method, select a theory or conceptual framework, create a coding scheme, generate hypotheses, prepare an analysis plan, or select a journal.

Differentiate the five proposals from one another as much as possible. Respond in Turkish.

**Table 1.** P0 Candidate Research Problems and Cross-System Semantic Convergence

| System | No. | Candidate problem direction | Primary semantic cluster |
|---|---|---|---|
| ChatGPT | #1 | Role and decision authority in human–AI collaboration | Role / decision authority |
| | #2 | Epistemic status of evidence, assumptions, and intuition | Epistemic calibration |
| | **#3** | **Construction of continuity and project memory** | **Continuity / memory** |
| | #4 | Early fixation, scope expansion, and reopening of decisions | Design decisions / fixation |
| | #5 | Transformation of a non-technical founder's production capacity | Technical capacity |
| Gemini | #1 | Deep personal disclosure and role shift | Personal disclosure / role shift |
| | #2 | Sycophancy and critical validation strategies | Epistemic calibration |
| | **#3** | **Context degradation and human-built cognitive scaffolding** | **Continuity / memory** |
| | #4 | AI as compensatory socio-technical capital | Technical capacity |
| | #5 | Epistemic authority and agency in design decisions | Design agency / authority |
| Claude | **#1** | **Production of continuity in stateless systems** | **Continuity / memory** |
| | #2 | Negotiation and management of the AI role | Role / decision authority |
| | #3 | Demand for honesty and epistemic trust calibration | Epistemic calibration |
| | #4 | Execution capacity of a non-technical founder | Technical capacity |
| | #5 | Contribution, ownership, and decision architecture in creative work | Design agency / ownership |

*Note. Each of the 15 candidate problems is represented once under its primary semantic cluster. Semantic affinities were not necessarily mutually exclusive. Continuity/memory represented the clearest direct three-way convergence across systems. The selected candidates are shown in bold. Secondary selection criteria were data fit, originality, and publishability.*

## P1 Scholarly-Development Prompt — Original Turkish

Sana, gerçek bir proje geliştirme sürecinde doğal olarak oluşmuş, anonimleştirilmiş boylamsal bir insan–yapay zekâ etkileşim corpus'u veriyorum.

Bu corpus üzerinden tam bir akademik araştırma ve yayınlanabilir bir araştırma makalesi geliştireceğiz.

Kilitli araştırma problemi şudur:

Uzun süreli insan–yapay zekâ işbirliğinde çalışma sürekliliği, sistemin bağlam ve hafıza sınırlılıkları karşısında kullanıcı tarafından hangi pratikler ve dışsal kaynaklar aracılığıyla oluşturulur ve sürdürülür?

Çalışmanın en az Scopus Q2 düzeyinde bir dergide yayımlanabilecek bilimsel niteliğe ulaşmasını hedefliyorum.

Araştırmanın teorik konumlandırmasını, araştırma sorularını, yöntemini, analiz yaklaşımını ve akademik geliştirme sürecini corpus'un özelliklerine ve araştırma problemine dayanarak oluştur.

Bu süreci tek seferde tamamlamaya çalışma. Öncelikle corpus'u ve araştırma problemini değerlendirerek çalışmaya nasıl başlamamız gerektiğini öner ve gerekçelendir. Bundan sonraki adımları etkileşim içinde birlikte geliştireceğiz.

Çalışma sürecindeki değerlendirme, tartışma ve karar iletişimini Türkçe yürüt. Ancak ortaya çıkaracağımız makale İngilizce olacaktır; makale için yazılan başlık, özet, anahtar kelimeler, araştırma sorularının nihai ifadeleri ve tüm makale bölümlerini akademik İngilizce yaz.

## English Translation

I am providing you with an anonymized longitudinal human–AI interaction corpus that arose naturally during a real project-development process.

Using this corpus, we will develop a complete academic study and a publishable research article.

The locked research problem is:

*In prolonged human–AI collaboration, through which practices and external resources is continuity of work produced and sustained by the user in the face of the system's context and memory limitations?*

I aim to develop the study to a standard suitable for submission to a Scopus-indexed journal ranked in Q2 or above.

Develop the study's theoretical positioning, research questions, method, analytical approach, and scholarly development process on the basis of the characteristics of the corpus and the research problem.

Do not try to complete this process in a single step. First, evaluate the corpus and the research problem, then propose and justify how we should begin the study. We will develop the subsequent steps together through interaction.

Conduct the evaluation, discussion, and decision-making communication during the research process in Turkish. However, the article we produce will be in English; write the article title, abstract, keywords, final formulations of the research questions, and all manuscript sections in academic English.